\documentclass{aa}  

\usepackage{graphicx}
\usepackage{txfonts}
\usepackage{hyperref}
    \hypersetup{colorlinks=true,linkcolor=blue,citecolor=blue,urlcolor=blue}
    \makeatletter
    \renewcommand*\aa@pageof{, page \thepage{} of \pageref*{LastPage}}
    \makeatother
\usepackage{algorithm} 
\usepackage{algpseudocode}
\usepackage{gensymb}
\usepackage{tabularx,ragged2e,booktabs,caption}
\usepackage{multirow}

\begin{document}

   \title{Improving Earth-like planet detection in radial velocity using deep learning}
   
   \titlerunning{DL-exoplanet}
   \authorrunning{Yinan Zhao et al.}

   \author{Yinan Zhao
          \inst{1}
          \and
          Xavier Dumusque\inst{1}
          \and
          Michael Cretignier\inst{2}
          \and
          Andrew Collier Cameron\inst{3,4}          
          \and
          David W. Latham\inst{5}
          \and
          Mercedes L\'opez-Morales\inst{5}  
          \and
          Michel Mayor\inst{1}
          \and  
          Alessandro Sozzetti\inst{6}
          \and            
          Rosario Cosentino\inst{7}
          \and   
          Isidro G\'omez-Vargas\inst{1}
          \and          
          Francesco Pepe\inst{1}
          \and
          Stephane Udry\inst{1}
          }

   \institute{Department of Astronomy of the University of Geneva, 51 chemin de Pegasi, 1290 Versoix, Switzerland
   \and
   Department of Physics, University of Oxford, OX13RH Oxford, UK
    \and
    SUPA School of Physics and Astronomy, University of St Andrews, North Haugh, St Andrews, KY16 9SS, UK
    \and
    Centre for Exoplanet Science, University of St Andrews, North Haugh, St Andrews, KY169SS, UK
    \and
    Center for Astrophysics | Harvard \& Smithsonian, 60 Garden Street, Cambridge, MA 02138, USA  
    \and
    INAF - Osservatorio Astrofisico di Torino, Via Osservatorio 20, I-10025 Pino Torinese, Italy
    \and    
    Fundaci\'on Galileo Galilei - INAF, Rambla J. A. F. Perez, 7, E-38712 S.C. Tenerife, Spain    
   \\
              \email{yinan.zhao@unige.ch}
             }

  \abstract
   {Many novel methods have been proposed to mitigate stellar activity for exoplanet detection as the presence of stellar activity in radial velocity (RV) measurements is the current major limitation. Unlike traditional methods that model stellar activity in the RV domain, more methods are moving in the direction of disentangling stellar activity at the spectral level. As deep neural networks have already been proven to be one of the most effective tools in data mining, in this work, we explore here their potential in the context of Earth-like planet detection in RV measurements.}
   {The goal of this paper is to present a novel convolutional neural network-based algorithm that efficiently models stellar activity signals at the spectral level, enhancing the detection of Earth-like planets.}
   {Based on the idea that the presence of planets can only produce a Doppler shift at the spectral level while the presence of stellar activity can introduce a variation in the profile of spectral lines (asymmetry and depth change), we trained a convolutional neural network to build the correlation between the change in the spectral line profile and the corresponding RV, full width at half maximum (FWHM) and bisector span (BIS) values derived from the classical cross-correlation function. }
   {This algorithm has been tested on three intensively observed stars: Alpha Centauri B (HD\,128621), Tau ceti (HD\,10700), and the Sun. By injecting simulated planetary signals at the spectral level, we demonstrate that our machine learning algorithm can achieve, for HD\,128621 and HD\,10700, a detection threshold of 0.5 m/s in semi-amplitude for planets with periods ranging from 10 to 300 days. This threshold would correspond to the detection of a $\sim$4\,$\mathrm{M}_{\oplus}$ in the habitable zone of those stars. On the HARPS-N solar dataset, thanks to significantly more data, our algorithm is even more efficient at mitigating stellar activity signals and can reach a threshold of 0.2 m/s, which would correspond to a 2.2\,$\mathrm{M}_{\oplus}$ planet on the orbit of the Earth.}
   {To the best of our knowledge, it is the first time that such low detection thresholds are reported for the Sun, but also for other stars, and therefore this highlights the efficiency of our convolutional neural network-based algorithm at mitigating stellar activity in RV measurements.}  

   \maketitle
%

%

\section{Introduction}

The detection of low-mass planets using the radial-velocity (RV) technique remains extremely challenging in the presence of stellar activity. Stellar activity is the result of complex underlying physical processes happening at different timescales \citep[e.g.,][]{Meunier-2010a,Meunier-2017aa}. For example, granulation and super granulation contribute to stellar activity most on timescales of hours to a few days \citep[e.g.,][]{Del-Moro-2004a, Del-Moro-2004b, Dumusque-2011aa}, while dark spots and bright faculae regions contribute to photometric and RV variations on the timescale of the host star’s rotational period \citep[e.g.,][]{Saar-1997ApJ, Dumusque-2014b, Zhao-2023A&A}. On the timescale of years, the magnetic cycle of the host stars contributes to stellar activity the most \cite[][]{Lovis-2011b, Haywood-2016MNRAS, Haywood-2022ApJ}.

Machine learning is one of the most efficient and successful tools to handle large amounts of data in the scientific field. Many algorithms based on machine learning have been proposed to mitigate stellar activity to better detect low-mass and/or long-period planets. These algorithms can be classified into two categories: supervised learning and unsupervised learning. The advantage of supervised learning is that the proposed model contains a large set of variables and has the ability to produce relatively accurate predictions based on the training data. The Gaussian process (GP), as a supervised learning algorithm, is widely used to mitigate stellar activity using photometric or spectroscopic data. By carefully selecting the parametric kernels and priors, GP can efficiently model stellar activity \cite[][]{Perger-2021aa,Nicholson-2022MNRAS,Aigrain-2023ARA&A}. The potential drawback of GP is data overfitting when the hyperparameters or priors of GP are not chosen wisely. This may lead to erroneous parameter estimates for the detected planets \citep[e.g.,][]{Blunt:2023aa} or to the difficulty of resolving weak planetary signals as they might be absorbed by the fitted GP \citep[e.g.,][]{Langellier:2020aa}. Some previous work has already demonstrated the power of the convolutional neural network (CNN) in detecting transiting exoplanet signals \cite[][]{Shallue-2018AJ,Dattilo-2019AJ}, and also in mitigating stellar activity in RVs. In \cite{Beurs-2022AJ}, the authors explain how they used a CNN to efficiently model stellar activity at the level of the cross-correlation functions \citep[CCF][]{Baranne-1996} derived from the HARPS-N solar observations \citep[][]{Dumusque-2015ApJ,Phillips:2016aa,Collier-Cameron-2019MNRAS,Dumusque-2021aa}. The results show that solar activity is significantly reduced and it was possible to use this framework to detect a planetary signal with a period of 365 days and an amplitude of 0.4 m/s. \cite{Perger-2023aa} used several stellar activity indices as inputs to train a CNN and predict the corresponding stellar activity affecting the RVs. The CNN application on $\epsilon$ Eridani and AU Microscopii showed that the stellar activity signal of the two stars can be greatly reduced. \cite{Nieto-2023aa} trained a convolutional neural network in the periodogram domain to classify whether the planetary signals can be detected or not in RV time series affected by correlated noise. The performance on the simulated RV data shows an improvement of $28\%$ fewer false positives. These successful applications demonstrate that CNNs are powerful tools to model stellar activity and enhance the detectability of planetary signals embedded into astrophysical noise; however, they also reveal the disadvantages of CNNs. Indeed, the major drawbacks of CNNs are the hyperparameter tuning and the generalization from the training set. There are many hyperparameters that need to be tuned during the training process. This part usually involves human interference and may not be optimal for different stellar types. Even when the tuning is done properly, the fact that a CNN model contains a large number of free variables makes it difficult to generalize the target features from the training data, which can lead to bad predictions.

Many methods based on unsupervised learning have also been proposed to mitigate stellar activity at either the CCF level or the spectral level. \cite{Davis-2017ApJ} developed a principle component analysis (PCA) algorithm for simulated stellar activity affected spectral time series. This algorithm showed that eigenvectors of PCA are greatly affected by spectral line shape variation. \citet{Cretignier:2020aa,Cretignier-2022aa} further explored this idea and converted stellar spectra into flux-flux gradient space and implemented PCA to mitigate stellar activities in HD10700 ($\tau$ Ceti) and HD12861 ($\alpha$ Cen B) RV datasets. On 5 years of HD12861 data, the authors show that stellar activity at the rotational period was significantly removed and the RV root mean square was significantly reduced from $2.44\, \rm{m/s}$ down to $1.73\, \rm{m/s}$. \cite{Collier-Cameron-2021MNRAS} applied PCA to the autocorrelation function (ACF) of HARPS-N solar data CCF and were able to disentangle stellar activity from a planet signal of semi-amplitude $\sim0.4\,\rm{m/s}$. Compared to supervised learning algorithms, unsupervised learning algorithms have few hyperparameters to be tuned and may avoid overfitting; however, they strongly suffer from the quality of the input data. For example, some low-rank eigenvectors derived from PCA may be very sensitive to the noise level or outliers in the input data \citep[e.g.,][]{Ould-Elhkim:2023aa}. This may lead to a challenge in selecting proper sets of latent features for stellar activity removal. Furthermore, the limitation of employing PCA lies in its linear decomposition of inputs. Given that stellar activity represents a complex physical process, important features or proxies may be embedded in spectral time series in a nonlinear manner, which may not be captured due to the linearity of inputs. CNN, in contrast, is known for its ability to capture nonlinear features in input data.

In this paper, we propose a CNN-based algorithm that has the ability to efficiently remove stellar activity and improve the detection efficiency of low-mass, long-period exoplanet signals. The CNN architecture and hyperparameters can be automatically optimized for different stellar types. We describe the preprocessing of the spectral data and the strategy to generate training data in Section \ref{sec2}. The architecture of the neural network and methods for automatic hyperparameter tuning are described in Section \ref{sec3}. In Section \ref{sec4}, we explain how we applied the CNN to three stars: the Sun, HD128612, and HD10700, and we derived the detection limits of low-mass exoplanets for these stars. Finally, we draw our conclusion in Section \ref{sec4}. The code of this algorithm is publicly available on GitHub \footnote{code available here \url{https://github.com/YinanZhao21/DeepLearningShell}}

\section{Spectra preprocessing and dataset preparation}\label{sec2}

Two recent studies have shown that a CNN framework can be applied to mitigate stellar activity in RV measurements. \citet{Perger-2023aa} uses as input of a CNN several time series (CCF FWHM, BIS and contrast, multiband photometry) to model as output the RV measured as the mean of a Gaussian fit to the CCF. Although the CCF, FWHM, BIS and contrast contain some information about the CCF asymmetry, they do not necessarily capture all the information for optimal stellar activity modeling. \citet{Beurs-2022AJ} chose to use as input the full CCF to model stellar activity on the HARPS-N solar data. However, as demonstrated in \citet{Dumusque-2018aa}, stellar activity can be significantly reduced by carefully selecting certain absorption lines in the spectra. Therefore, there is clearly some information about stellar activity at the spectral level that a CCF might not be sensitive to, as by construction, the CCF is an average of most of the lines in a spectrum. If spectral lines are affected differently by stellar activity, which should be the case as lines are formed at different depths in the stellar photosphere and are therefore affected differently by the magnetic field, a CCF will wipe out those differential effects and thus it might be more difficult to probe and model stellar activity at the CCF level than at the spectral level. Some pioneering works described in \citet{Cretignier:2020aa}, \citet{Wise:2022aa} and \citet{AlMoulla:2022aa} demonstrate how spectral lines are differently affected by stellar activity, and explain the observed behavior based on the physics of stellar atmospheres. 

The above discussion strongly motivates the development of methods to model stellar activity at the spectral level and there are already a few frameworks that show promising results \citep[e.g.,][]{Binnenfeld-2020aa,Binnenfeld-2022aa,Shahaf-2023MNRAS}. However, a HARPS-N spectrum has hundreds of thousands of data points. This is three orders of magnitude larger than the input data used in the CNN frameworks described in the preceding paragraph, and using spectra as input of a CNN will very significantly increase the complexity of the CNN architecture, with a risk of overfitting at the end, without speaking about the time needed to train the network. Our approach is therefore to use a dimensional space which is in-between the stellar spectrum and a CCF in terms of input dimension and in terms of capturing differential effects due to stellar activity between spectral lines. The chosen space is the normalized flux-flux gradient space (hereafter shell spectral representation) described in \citet{Cretignier-2022aa}. This space has been specifically designed to reduce the dimension of a spectrum while maximizing the information induced by the inhibition of the convective blueshift, the main contributor of stellar activity on solar-type stars. A detailed description of the transformation from spectrum to shell is presented in Section \ref{sec2.1}. 

\begin{figure}[htbp]
  \includegraphics[scale=0.3]{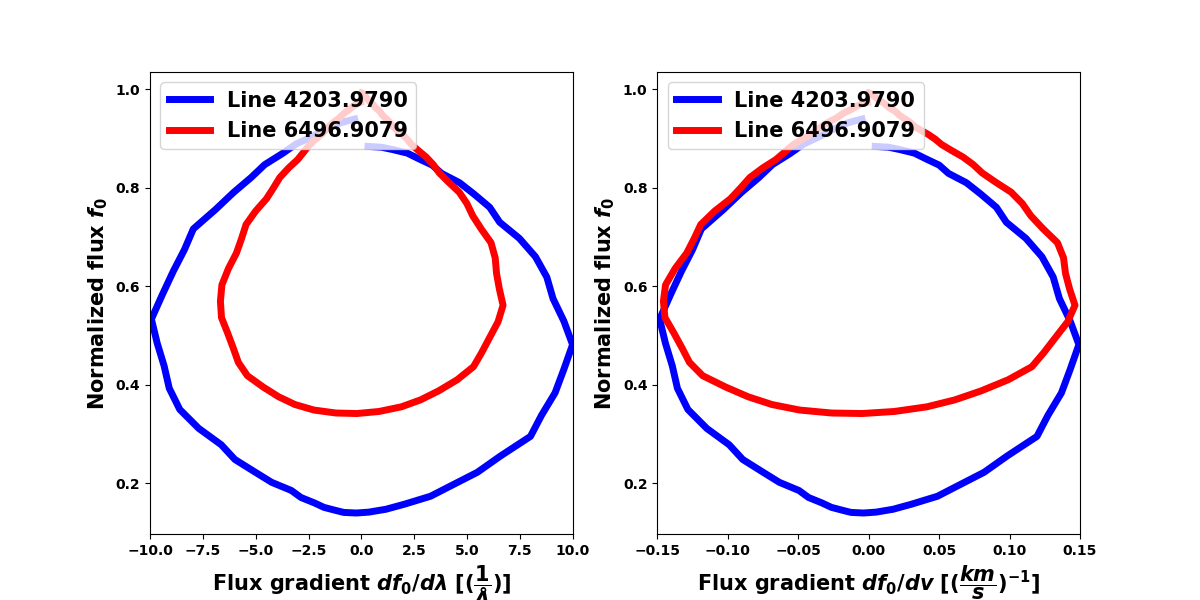}
  \caption{Shell spectral representation of two spectral lines at opposite edges of the visible spectral range. \emph{Left:} Two spectral lines in the $(f_{0}, \frac{df_{0}}{d \lambda} )$ shell space. Line $4203.9790 \AA$ at the blue part of the spectrum has the larger loop than Line $6496.9079 \AA$ at the red part of the spectrum in  $(f_{0}, \frac{df_{0}}{d \lambda} )$ space. \emph{Right:} Line $4203.9790 \AA$ and $6496.9079 \AA$ in the $(f_{0}, \frac{df_{0}}{d v} )$ shell space. The line in the blue part of the spectrum follow the same loop as the line in the red part when considering the common normalized flux, from 0.9 to 0.4. Therefore, the chromatic effect due to dispersion is suppresed in the $(f_{0}, \frac{df_{0}}{d v} )$ shell space.}
  \label{Fig_chromatic}%
  \end{figure}

The bottleneck in training a neural network to mitigate stellar activity is the limited number of observed spectra. Stars significantly observed have at most a few thousand observations, and it is expected that stellar activity behavior changes for different spectral types, therefore it is dangerous to merge the data of different stars together. It is tempting to rely on simulated data to train a CNN, however, due to the complexity of the physics behind stellar activity, good performance on simulated data usually leads to a moderate performance on the real observed stellar spectra. In order to solve the limitation of the training sample size, \cite{Beurs-2022AJ} first introduced the cross-validation technique to increase the sample size during the training process. In this work, we adopt this technique to train our neural network for different stars. We describe this in Section \ref{sec2.2}.

  \begin{figure*}[htbp]
  \includegraphics[scale=0.4]{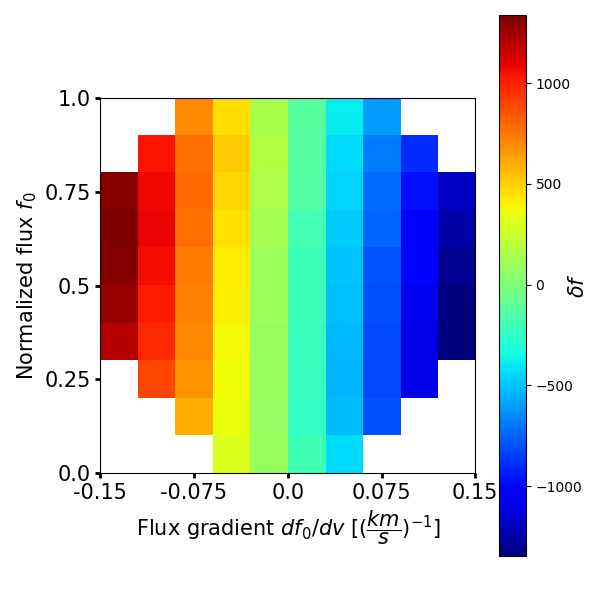}
  \includegraphics[scale=0.4]{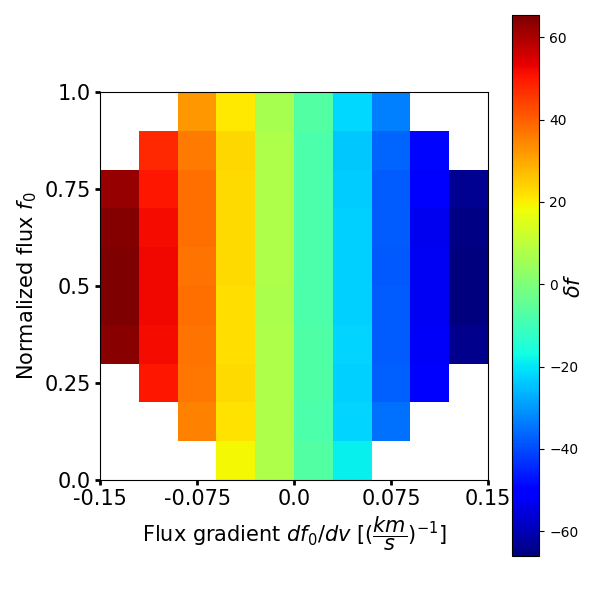} 
  \includegraphics[scale=0.4]{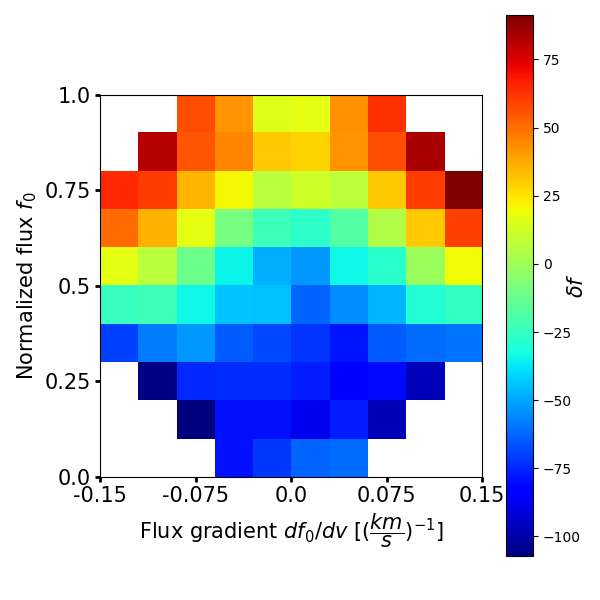}
  \caption{Examples of solar spectral shells in $(f_{0}, \frac{df_{0}}{d v} )$ shell space. \emph{Left:} An original spectrum from HARPS-N observation is transformed into the $(f_{0}, \frac{df_{0}}{d v} )$ shell space. \emph{Middle:} Solar spectral shell due to a Doppler effect. The Doppler shell is obtained by injecting a $5\,\rm{m/s}$ Doppler shift into the HARPS-N solar master spectrum $f_{0}$. \emph{Right:} Solar spectral shell due to spectral shape change. We derived the shape shell by projecting the original shell onto the Doppler shell and subtracting the component due to Doppler effect from the original shell.}
  \label{Fig_shell_example}%
  \end{figure*} 
  
\subsection{Definition and motivation of the shell spectral representation} \label{sec2.1}

The shell spectral representation was first defined in \citet{Cretignier-2022aa} and in this section, we revisit its definition to be more sensitive to stellar activity. From the physics of stellar atmospheres, all photospheric spectral lines share the same bisector, which traces the velocity of stellar convection as a function of depth into the photosphere \citep[e.g. ][]{Dravins:2008aa, Gray-2009}. The cores of shallow spectral lines, probing deep regions into the photosphere, are strongly blueshifted compared to the laboratory wavelength of the line transitions, which is not the case for the cores of strong lines, which are formed toward the top of the photosphere \citep[e.g.][]{Reiners:2016aa, AlMoulla:2022aa}. Therefore, convection is strong deep inside the photosphere, with values of a few hundred of m/s for the solar case, and nearly zero close to the chromosphere. The strong magnetic fields present inside active regions tend to inhibit stellar convection \citep[e.g.][]{Cavallini-1985}, and therefore the cores of shallow lines, but also the wings of strong lines close to the continuum will be strongly affected by this inhibition of convective blueshift, while the core of strong lines should not be affected as there is no convection to inhibit. In summary, the bisector of spectral lines encodes a lot of information regarding stellar activity. By cross-correlating a stellar spectrum with a mask that has different weights for each spectral line, as it is generally done to optimize RV precision \citep[][]{Pepe-2002a}, the cross-correlation technique destroys the original line bisector and some information about stellar activity is lost in the process. To be more sensitive to stellar activity, we should consider a CCF without line weighting, or the ``shell'' spectral representation, described in \citet{Cretignier-2022aa} and revisited below.

Let us consider two spectra $f(\lambda)$ and $f_0(\lambda)$ that are defined on the same grid in wavelength $\lambda$ but that are Doppler-shifted by an amount that is small compared to the sampling\footnote{typical sampling for a high-resolution $R=100'000$ spectrograph is several hundreds of m/s, so this assumption is most of the time valid when working with high-precision RVs}. The observed flux variation at pixel $i$ can be expressed as \citep[see also][]{Bouchy-2001}:
    \begin{equation} \label{eq:1}
    f(\lambda_i) - f_{0}(\lambda_i) = \frac{d f_{0}(\lambda_i)}{d \lambda_i} \cdot \delta \lambda_i = \frac{d f_{0}(\lambda_i)}{d \lambda_i} \frac{\delta V_i}{c} \lambda_i,
      \end{equation}
considering that the Doppler shift formula is $\delta V/c = \delta \lambda/\lambda$, $\delta V$ being the Doppler shift and $c$ the speed of light. We can therefore rewrite:
   \begin{equation} \label{eq:2}
    \delta V_i = \frac{c} {\lambda_i} \cdot (f(\lambda_i) - f_{0}(\lambda_i))  \cdot \left(\frac{d f_{0}(\lambda_i)}{d \lambda_i}\right)^{-1} = \frac{c} {\lambda_i} \cdot \delta f_i \cdot \left(\frac{d f_{0}(\lambda_i)}{d \lambda_i}\right)^{-1}.
      \end{equation}
Following this equation, \citet{Cretignier-2022aa} define the shell spectral representation as being normalised flux as a function of flux derivative with respect to wavelength, $\left(f_0,\frac{d f_{0}}{d \lambda}\right)$. With this convention, two similar spectral lines in the blue and red parts of the spectrum will follow different loops in the shell spectral representation due to dispersion (see left panel of Fig. \ref{Fig_chromatic}). The way line bisector, or velocity of convection as a function of depth, is therefore chromatic in this definition of the shell spectral representation, which will alter the signal induced by stellar activity similarly to a weighted CCF. To prevent that and optimize the extraction of stellar activity, we propose in this paper to define the shell spectral representation as being normalised flux as a function of flux derivative with respect to velocity, $\left(f_0,\frac{d f_{0}}{dv}\right)$. In this new shell spectral representation, rewriting Equation~\ref{eq:2}:
   \begin{equation} \label{eq:3}
    \delta V_i = \delta f_i \cdot \left(\frac{d f_{0}(v_i)}{dv_i}\right)^{-1},
      \end{equation}
we easily see that a pure Doppler shift is simply a gradient in the $x$ direction (see middle panel of Fig.~\ref{Fig_shell_example}). From now on, the term spectral shell will be used to describe the shell spectral representation in velocity space, $\left(f_0,\frac{d f_{0}}{dv}\right)$.

Another difference with the spectral shell we use in this paper compared to what is published in \citet{Cretignier-2022aa} is the part of the spectrum that is used to build the spectral shell. In this previous work, the authors were using the full spectrum, which can be a disadvantage as other effects than stellar activity and planetary signals might contaminate the spectral shell. While in a weighted CCF, the weight, which is generally proportional to the contrast of spectral lines, strongly down weight very shallow spectral lines that contain little information, this is not the case in a spectral shell as no weight is applied. To prevent contamination, we therefore only measure the spectral shell using lines that are not blended and that are deeper than $0.1$ \citep[see line selection described in][]{Cretignier:2020aa}. To increase the signal to noise (S/N) of the spectral shell, we bin it evenly down to a dimension of $10\times 10$. This strongly increases the S/N at the edges of the spectral shell and prevents outliers in those regions which could strongly hamper our neural network approach described in Section.~\ref{sec3}. An example of a spectral shell derived from our subset of lines is shown in the left panel of Fig \ref{Fig_shell_example}. 

The main idea of this work is try to link the shape variations of spectral shell due to stellar activity with the corresponding RV values, being very careful that if a planetary signal is present in the original data, this one is not modeled by our neural network \footnote{We note that we consider a regime where the planetary signal is much smaller than the signal due to activity, and therefore the RVs that are used to train our neural network are dominated by the signal from stellar activity.}. We must therefore remove from the spectral shell any component that corresponds to a Doppler shift in order to obtain what we refer to as a shape shell. This can be done by first estimating the Doppler shell $S_{Doppler}$, which can be derived by injecting an arbitrary Doppler shift into the template spectrum $f_{0}$. Once this is done, the shape shell at time $t$, $S_{shape}(t)$ can be obtained from the spectral shell at the same time $S(t)$ using the formula:
\begin{equation} \label{eq:4}
S_{shape}(t) = S(t) - (S(t) \cdot S_{Doppler}) S_{Doppler}.
\end{equation}
An example of shape shell (Doppler free spectral shell) is shown in the right panel of Fig \ref{Fig_shell_example}.

\subsection{Spectral data as cross-validation dataset}\label{sec2.2} 

In the application of deep learning, the preparation of training data and test data is important for the success of solving any task. In the context of this work, the main challenge lies in the amount of data that can be used for different types of stars. In the 6-year HARPS-N solar observations, there are $\sim30000$ hourly binned spectral data. In order to minimize the effect of granulation and super granulation, and other short-term effects, we daily binned the spectra \citep[e.g.][]{Dumusque-2011aa}. As a result, only $\sim1800$ daily binned spectra can be used to train and test different deep learning architectures. Speaking of stars other than the Sun, there are $\sim400$ daily binned spectra for HD10700 ($\tau$ ceti) and $\sim300$ daily binned spectra for HD128621 ($\alpha$ CenB) in the HARPS archive dataset. The limitation of the available data for different stars is therefore the main obstacle. One possible solution would be to use simulated spectra as the training sets for different stellar types and apply the trained neural network to the observed spectra. Although the recently released SOAP-GPU is able to simulate stellar activity at spectral level for stellar type from F9 to K2 \citep{Zhao-2023A&A}, stellar activity is a complex process and although most of the main physical processes are included in the code, the amplitude and covariance of those processes are based on either solar observations or numerical simulations. In addition, instrumental systematics and telluric contamination are not included in SOAP-GPU, and the presence of those in real observations likely makes it more difficult for a neural network to converge to what we want, modeling only the stellar activity component. Preliminary tests showed that the difference between SOAP-GPU simulated spectra and observed spectra is significant and although SOAP-GPU data were used at the start to develop and test different neural network architectures in an idealistic case, we quickly moved to real data as a trained neural network on the simulated data was not delivering good results when applied to real observations.

To augment the dataset based on real observations, we can use cross-validation as a resampling method. In general, cross-validation splits the data into training and test sets for different realization. For each realization, the training and test sets are used to train a machine learning model. The final result is the ensemble of the trained models for k iterations. \cite{Beurs-2022AJ} first used this method to handle the limited solar spectra of HARPS-N observation when training the neural network to model solar activity at CCF level. Here we also use this method to train neural network for different stars. Throughout this paper we use $k=10$, which is the number of realization in which the input data are split into. In details, the spectral shells are randomly divided into 10 groups, For each iteration, one group is designated as the test set, while the remaining groups are used to train a model. The final results is the summary of the evaluations conducted on the test sets across all iterations. The process of the cross-validation is summarized in the pseudo-code in \textbf{Algorithm \ref{algorithm1}}.

\begin{algorithm}
        \caption{Cross-validation} 
        \begin{algorithmic}[1]
                \For {$k=1,2,\ldots 10$}                
                    \State Take $90 \%$ of the input spectra as the training set.
                    \State Take the remaining  $10 \%$ of the input spectra as the test set.
                    \State Train neural network on the training set and evaluate it on the test set.
                    \State Save the results of the test set.
                \EndFor
        \State Summarize all the results.      
        \end{algorithmic} 
\label{algorithm1}        
\end{algorithm}

\begin{figure}[htb]
\centering
\includegraphics[scale=0.45]{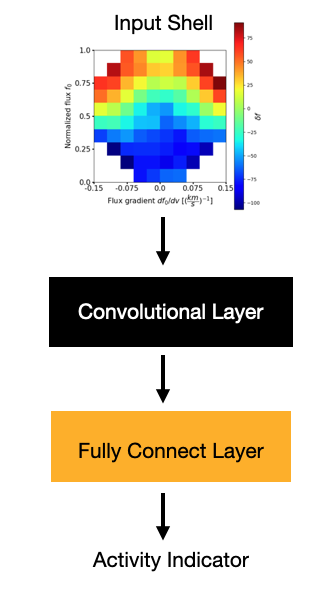}
\caption{Demonstration of a neural network architecture to be optimized by $Optuna$. Shape shell with dimension of $10\times 10$ is feed into the neural network. The neural network consists of convolutional blocks, an adaptive layer and fully connected layers. In order to avoid overfitting planetary signals in RV time series, the calcium activity indicator $\log(R'_{HK})$ is used as output of the neural network to search for its best architecture.}
\label{NN_indicator}%
\end{figure}

\section{Deep neural network model}\label{sec3}

We aim to build a neural network that can predict the RV induced by stellar activity from spectral shape shells in the $(f_{0}, \frac{df_{0}}{d v} )$ space. If this works, and if our network is insensitive to planetary signals, then simply removing the predicted RVs from the raw RVs should enhance the significance of planet signals embedded into stellar signals. To prevent absorbing planetary signals, we first work with the shape shells, which are free from any Doppler component, as the input of our neural network. In addition, rather than predicting only the RVs, which are derived from the CCF technique and therefore contain tiny planetary signals that the neural network could tend to model, we also predict simultaneously to the RVs, the FWHM and BIS derived from the CCF, similarly to what is done in multidimensional GP regression to prevent planetary signal overfitting \citep[e.g.][]{Rajpaul-2015MNRAS, Barragan-2022MNRAS, Delisle:2022aa}. For stars where planetary signals have already been detected, we remove the known planetary signals before feeding the spectral time series and the corresponding RV time series into the deep learning framework.

Since shape shells are two-dimensional data, we use a convolutional neural network to achieve this regression task. The deep learning model is implemented using PyTorch \citep{paszke2017automatic}. Regarding the challenging task of hyperparameter tuning, which is generally done manually, it is unrealistic here, when considering several stars, to each time go through this manual process. We therefore implemented an automatic hyperparameter tuning algorithm to choose the best neural network architecture for different input stars. We describe the automatic neural network initialization in Section. \ref{3.1} and the deep learning algorithm to model stellar activity from shape shell in Section. \ref{3.2}.

\subsection{Neural network initialization for different stars}\label{3.1}

The deep neural network used in this study has three basic components: an input layer, hidden layers and an output layer. In this case, the input layer is the input shape shell with a dimension of $10\times 10$. Although a higher resolution of the input shell includes more complex features, it also leads to a S/N decrease, mainly at the edges of the shell. Considering that most of the stars have only limited data, it is recommended to privilege high S/N input data rather than low S/N input data with more complexity. 

Three blocks are used in the hidden layer: convolutional layers (hereafter Conv layer), pooling layers and fully connected layers (hereafter FC layer). As the key element in convolutional neural network architecture, each convolutional layer contains number of length-fixed nodes called filters. Each filter convolves the input shell with a set of weights \citep{Krizhevsky-NIPS2012}. The weights in each filter is optimized by the back propagation algorithm using cost function \citep{Rumelhart-1986nature}. The convolved image in each node is called a feature map. We denote the feature map from the $k$th node as $F^{k}$, whose filters are determined by the weights $W^{k}$ and bias $b^{k}$. The feature map can be obtained by: 
\begin{equation} \label{eq:5}
F^{k} =  (W^{k} \times S_{shape, T}) + b^{k}.
\end{equation} 
Important features in the input shell is maximized by this process and will be passed to the next layer.

A pooling layer is used to reduce the spatial size of the feature maps from previous convolutional layers while important features in feature maps are preserved by calculating the average or max value when the filters of pooling layer convolve with feature maps. In this work, we use the adaptive average pooling layer after the convolutional blocks and before passing the flatten vectors into FC layers. The advantage of adaptive pooling is to have better spatial adaptability than the traditional average or max pooling which has a fixed window size \citep{Stergiou-2021arxiv}.

Once the feature map after the pooling layer is flattened, it is connected to FC layers. Each element in the flattened feature map is connected to neurons in the FC layers, which perform a linear transformation by using a weight matrix. After each FC layer, we implement a dropout layer which will randomly drop out neurons with a specified ratio, as a measure to alleviate overfitting during the training process \citep{Hinton-2012arxiv}.
 
   \begin{table}[H]
   \captionof{table}{Hyperparameters to be optimized by $Optuna$.} 
        \resizebox{\columnwidth}{!}{\begin{tabular}{c c c}
            \hline
            Parameters     &  Value:min, max  &  step size  \\
            \hline
            No. convolutional layers &  2, 3  &  1  \\
            No. convolutional filter in each layer  & 16, 128 & 16\\
            Filter size of first layer  & 1, 5 & 2 \\
            Filter size of other layers  & 1, 3 & 2 \\            
            No. FC layer &  2,3  & 1  \\
            No. FC neurons in each layer   & 32, 512 & 32\\
            Dropout ratio in each FC layer & 0.2, 0.5 & 0.1 \\
           \hline            
            Optimizer & Adam, RMSprop, SGD & \\
            \hline
         \end{tabular}}
      \label{tab:my_label}
   \end{table}

Since we want to use the neural network to predict radial velocity corresponding to stellar activity from shape shell, the Mean Squared Error (MSE) function is used as the cost function. During the training process, we minimize the cost function:
\begin{equation} \label{eq:6}
C = \frac{1}{N}\sum^{N}_{i=1}\left[RV_{pred}-RV_{CCF}\right]^2,
\end{equation} 
where $N$ is the number of input shape shell during the training. $RV_{CCF}$ is the ground truth RVs measured from spectral CCF and $RV_{pred}$ is the predicted RV based on the input shape shell.

\begin{figure}[htbp]
\centering
  \includegraphics[scale=0.35]{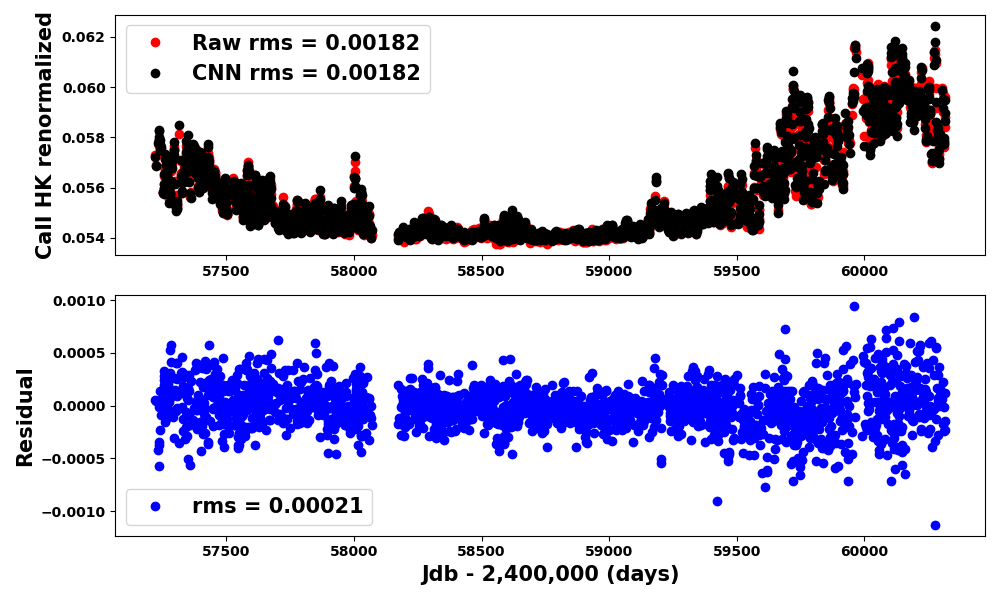}
  \caption{Results of stellar activity modeling on the calcium activity index derived from the HARPS-N solar data with our trained neural network. Top: 
  The calcium activity index time series derived from HARPS-N solar spectra is labeled in red, while the predicted calcium activity index time series, obtained using the neural network optimized by $Optuna$, is labeled in black. Bottom: The residual time series of the calcium activity index.}
  \label{Fig_activity_optuna}%
  \end{figure}

In order to find the best neural network architecture that is appropriate given the input data of a given star, we use the automatic hyperparameter optimization framework $Optuna$ \citep{Akiba-2019arXiv} to search for the most appropriate hyperparameters of the neural network. Since the input shape shell only has a dimension of $10\times 10$, a very deep neural network may lead to data overfitting. Therefore, the maximum numbers of both convolutional and FC layers are set to three. We set the maximum filter size of the first convolutional layer to five, in order to have a large receptive field, while the maximum filter size of the remaining layers is set to three. The last hyperparameter that we need to tailor is the optimizer algorithms used during the training. The possible options are Adam \citep{Kingma-2014arXiv}, RMSprop and Stochastic Gradient Descent (SGD). The number of trials for a given star is set to 50. The search space of all the hyperparameters is summarized in Table \ref{tab:my_label}.

\begin{figure}[htp]
\centering
\includegraphics[scale=0.35]{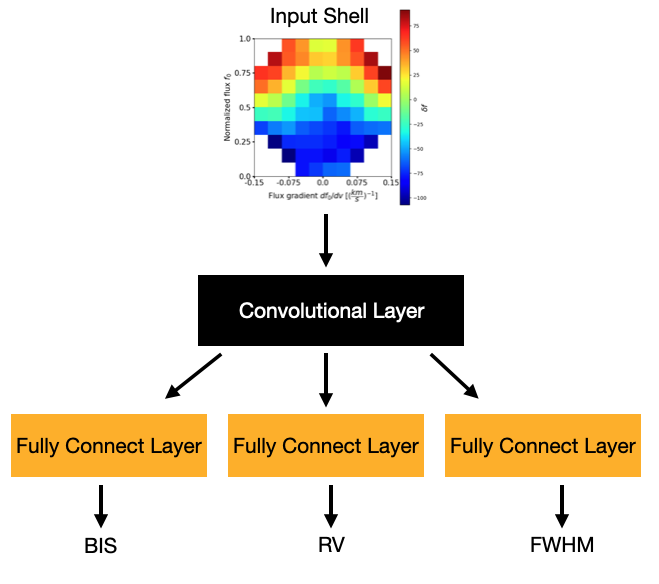}
\caption{Demonstration of a neural network architecture to predict the RV, FWHM, and BIS components of stellar activity. The FC blocks connected to each output have the same architecture and share the same feature maps from the convolutional blocks.}
\label{NN_multi_task}%
\end{figure}

Since the idea of this work is to use the neural network to predict RVs corresponding to stellar activity from spectral shape shell, the main obstacle we have to handle during optimizing the neural network architecture is the presence of planetary signals in the RV time series. For most of the stars other than the Sun, we have no prior information about the presence of planetary signals. If we optimize the neural network architecture by minimizing the MSE equation (see Eq.~\ref{eq:6}), the neural network is very likely to fit not only the RV component of stellar activity but also absorb the RV component of possible planetary signals. 

In order to avoid this issue when finding the best neural network architecture, we use a measured activity index as the output of our neural network. This optimizes its architecture and loosely constrains the neural network to prevent it from fitting potential planetary signals in the RV spectral time series. Specifically, we use the calcium activity index $\log(R'_{HK})$ for optimization as this indicator is derived from the S-index, primarily containing information about solar plage \citep{Cretignier(2024)} and does not contain any planetary information. Given that we want the trained CNN to predict the RVs, FWHM and BIS simultaneously, it is wise to use activity indicators other than FWHM or BIS to optimize the neural network hyperparameters. Relying on the optimization by FWHM and BIS for CNN hyperparameters could result in an overweighting of FWHM and BIS, potentially leading to an imbalance in the multi-task predictions, with reduced emphasis on RVs. 

If the architecture is chosen without considering any planetary signal, it is likely to absorb less planetary signal when trained on RV data. Therefore, instead of considering Eq.~\ref{eq:6} to optimize our neural network, we use as cost function:
\begin{equation} \label{eq:7}
C_{optuna} = \frac{1}{N}\sum^{N}_{i=1}\left[\log(R'_{HK,\,pred})-\log(R'_{HK,\,obs})\right]^2.
\end{equation} 
The basic architecture to be optimized by $Optuna$ is shown in Fig.~\ref{NN_indicator}.

$Optuna$ monitors Eq. \ref{eq:7} for each trial of hyperparameters and selects the optimal set with minimum $C_{optuna}$. Throughout optimization, we maintained a fixed training epoch of 300 steps per trial. To prevent overfitting during the optimization, the algorithm outputs the minimum $C_{optuna}$ value, rather than the $C_{optuna}$ value at the final epoch, for each trial. To ensure convergence within the fixed training epochs, we employed input shell data derived by injecting the YARARA stellar activity correction component back into the YARARA-cleaned spectra (for detailed spectrum processing descriptions, see Section. \ref{sec4}). An example of the neural network, featuring the best hyperparameters derived from $Optuna$, applied on the activity indicator time series is illustrated in Fig. \ref{Fig_activity_optuna}.
 
\subsection{Neural network application in stellar activity mitigation}\label{3.2}

Once the hyperparameters of the neural network architecture has been optimized to the given data, we start training our neural network. To further prevent our neural network to fit potential planetary signals in the RV time series, we ask our neural network to predict simultaneously the RVs, FWHM and BIS time series. The ground truth of those values are derived from the CCFs which is obtained by using the specific line mask mentioned in Section \ref{sec2.1}. Since we have three outputs from the neural network, the cost function we want to minimize is :
\begin{equation} \label{eq:11}
C_{total} = W_{RV}C_{RV} + W_{FWHM}C_{FWHM} + W_{BIS}C_{BIS},
\end{equation} 
where
\begin{equation} \label{eq:8}
C_{RV} = \frac{1}{N}\sum^{N}_{i=1}\left[RV_{pred}-RV_{CCF}\right]^2,
\end{equation} 

\begin{equation} \label{eq:9}
C_{FWHM} = \frac{1}{N}\sum^{N}_{i=1}\left[FWHM_{pred}-FWHM_{CCF}\right]^2,
\end{equation} 

\begin{equation} \label{eq:10}
C_{BIS} = \frac{1}{N}\sum^{N}_{i=1}\left[BIS_{pred}-BIS_{CCF}\right]^2,
\end{equation} 
and where $W$ is the corresponding weight for the cost function of each output. Generally it is possible to derive the weight $W$ for each output during training \citep{Kendall-2017arXiv}. However, considering the limited data for each star, such an approach might not give good results. For simplicity, and to prevent planetary signal overfitting, we assign an equal weight of one for each output. 

Following some tests, we found it was difficult for the neural network to predict three outputs by only changing the output layer from 1 to 3. We therefore modified the neural network so that each output is connected to its own FC block. The three FC blocks have the same hyperparameters derived from the $Optuna$ optimization. The three individual FC blocks share the same feature maps derived from the convolutional blocks. The architecture of the used neural network to predict stellar activity is shown in Fig \ref{NN_multi_task}.  

The strategy outlined in Section \ref{3.1} and \ref{3.2} serves only as a streamlined approach to avoid of fitting potential planetary signals in the RV time series. Directly performing optimization and training the neural network using RVs, FWHM and BIS is rather challenging. Attempting optimization on the multi-task neural network by monitoring $C_{total}$ from Eq. \ref{eq:11} may result in underfitting some outputs. For instance, although the optimal set may yield the minimum $C_{total}$ with a fixed training epoch, its corresponding $C_{RV}$, $C_{FWHM}$, and $C_{BIS}$ may not exhibit a monotonically decreasing trend. When applying the neural network to stars with little stellar activities, underfitting of RV time series often occurs due to the dominance of instrumental signals in the input spectral shell.

\section{Model performance and application}\label{sec4}

In order to test our deep learning framework for different regimes of stellar activity, we trained and tested our neural network on three stars: the Sun, HD128621 (Alpha Centauri B) and HD10700 (Tau ceti). These stars are intensively observed by HARPS-N or HARPS and have enough spectra to train our neural network. The Sun serves as the golden target to test the performance of a deep learning framework on solar-like stellar activity with a nearly perfect sampling. HD128621 is more active and can be used to evaluate the performance of our framework when stellar activity is stronger. HD10700 seems extremely quiet in RV, likely due to its pole-on configuration \citep[][]{Korolik:2023aa}, and therefore can test our framework when no or very little stellar activity is observed. Results on this star can also show the influence of instrumental and other sources of noise on the performance of the neural network. All the spectra of these stars are preprocessed by the YARARA data pipeline \citep{Cretignier:2021aa} to remove systemic errors such as induced by tellurics, detector stitching, ghosts and interference patterns. Another YARARA processing corrects for stellar activity seen at the spectral flux level and should mitigate the flux effect induced by active regions. 

In the initial version of the pipeline, stellar activity was corrected on the residual spectra time-series by a linear decorrelation with the S-index. This is justified since the S-index is largely dominated by the contribution of plages at least on the Sun \citep{Cretignier(2024)}. This recipe was then slightly improved afterward. The stellar activity correction is now performed in the rest frame of the active regions at the stellar surface (Cretignier in prep.) by following the recommendations of \citet{Thompson(2017)}. A similar philosophy was also recently presented by \citet{Shahaf-2023MNRAS} where the authors assumed that the spectra of active regions were "known," while in our case we used the S-index as an empirical proxy for them.

Our deep learning framework is applied on the shape shells derived after all those YARARA corrections. We note that we will also investigate, in the subsections relative to each star, the effect of the different YARARA corrections on the performance of our neural network. For reader clarity, we adhere to the naming conventions for spectral products established in \citep{Cretignier-2023aa}. These include:
\begin{itemize} 
    \item YARARA products corrected of all spectral-level corrections, denoted as YV1.
    \item YV1 with reinjected activity, labeled as YVA.    
    \item YARARA products not corrected but continuum normalized, referred to as YV0.
\end{itemize}

\subsection{The Sun}

We used the solar spectra observed by the HARPS-N solar telescope ranging from 2015 July 19th to 2024 January 8th \citep[][Dumusque et al. in prep.]{Dumusque-2021aa}. There are 2040 daily binned spectra in total. The hyperparameters of the neural network architecture are derived by optimizing training our neural network to predict the $\log(R'_{HK})$ index of the Sun with $Otpuna$. The best neural network architecture for the Sun is summarized in Table \ref{tab:NN_sun}. We trained the neural network with those 2040 spectra by using \textbf{Algorithm \ref{algorithm1}} for 100 training epochs. The predicted RV, FWHM, and BIS span time series from the trained neural network are illustrated in Fig \ref{Sun_mad}.

\begin{table}[H]
\centering
\caption{Best neural network architecture for the Sun.}
\begin{tabular}{ c c c c c c}
                               \toprule

    \multicolumn{5}{c}{Input layer} \\ \midrule
    \multicolumn{5}{c}{Conv1 (128, 3)} \\
    \multicolumn{5}{c}{Conv2 (32, 1)} \\
    \multicolumn{5}{c}{Conv3 (48, 1)} \\ \midrule
    
    \multicolumn{5}{c}{Adaptive average pooling(3,3)} \\ \midrule
     
   FC1 (352) & \multicolumn{3}{c}{FC1 (352)} & FC1 (352) \\
   Dropout1 (0.2) & \multicolumn{3}{c}{Dropout1 (0.2)} & Dropout1 (0.2) \\

    FC1 (416) & \multicolumn{3}{c}{FC1 (416)} & FC1 (416) \\
 \midrule
     
   FWHM & \multicolumn{3}{c}{RV} & BIS \\  \bottomrule
\end{tabular}
\tablefoot{Adam algorithm is employed as the optimizer.}
\label{tab:NN_sun}
\end{table}

In the RV space, we found that the neural network produces a good fit of the RV time series. In the periodogram, we also found that the neural network can well capture the activity signal at half the solar rotational, which was difficult to model when using principal component analysis decomposition of spectral shell \citep[at least on HD128621, see][]{Cretignier-2022aa}. However, there is still some signal at the solar rotation that is not fully modeled. The root mean square (RMS) of the RV time series is reduced from $0.822\,\rm{m/s}$ to $0.663\,\rm{m/s}$. There are still some features in the RV residual time series. For example, two spikes are present around 58500 and 59000. This is likely due to instrumental systematic errors induced by a HARPS-N detector warm-up and by the change of the main ThAr lamp, respectively, which lead to errors in wavelength calibration. Those strong departures from the mean RV are likely responsible for the strong signals near a thousand days that appear in the periodogram of the RV residuals. Besides those strong features, the periodogram of the RV residuals up to 200 days is very clean except probably for a small stellar activity residual signal appearing at half the solar rotation. A level of $0.663\,\rm{m/s}$ in the residuals is comparable to what is predicted from supergranulation \citep[e.g. $0.86$ and $0.68\,\rm{m/s}$ in][respectively]{Lakeland:2024aa, AlMoulla:2023aa} and thus the obtained RV residuals might be limited by this source of the stellar signal that our neural network is not designed to model. In the FWHM space, the neural network can significantly reduce the RMS of the FWHM time series from $0.682\,\rm{m/s}$ to $0.553\,\rm{m/s}$. The activity signals at the solar rotation period and half of it is also well modeled. In the periodogram of the FWHM residuals, the only significant signals left are associated with one year and half of it. This is probably some residuals left from a bad correction of the yearly change in FWHM induced by a variation in the viewing angle of the Sun along the Earth orbit \citep[see Section. 3.2 in][]{Collier-Cameron-2019MNRAS}. In the BIS space, the activity signals seen at the rotation period of the Sun and half of it are also well modelled by the trained neural network. The residuals of the BIS derived from the CCFs are reduced from $0.480\,\rm{m/s}$ to $0.314\,\rm{m/s}$. In the periodogram of the bisector span residuals, we only see a significant signal at a very long period, which is of unknown origin.

We further tested the performance of our deep learning framework on the solar spectra which have no YARARA stellar activity correction. We injected the YARARA stellar activity correction component back into the YARARA-cleaned spectra before deriving the shape shells used to train our neural network. The results of the trained neural network are shown in Fig \ref{Sun_activity}. By injecting the YARARA stellar activity correction component back, we can clearly see the signal induced by the magnetic cycles in the RV and BIS time series. The trained neural network can well model this extra long-term trend compared to the YV1 spectral data analysis presented in the preceding paragraph and can significantly reduce the RMS of the RV time series from $2.206\,\rm{m/s}$ to $0.872\,\rm{m/s}$. The FWHM and BIS time series are fitted equally well as in the YV1 spectral dataset case. Besides the small increase of the RV RMS in the residuals compared to the YV1 spectral case, we can say that our neural network approach performs similarly on data that are not stellar activity cleaned with YARARA.

Finally, we tested the performance of the deep learning framework on the solar spectra in the case with no YARARA correction. Before training our neural network, we injected all the YARARA correction components back into the solar spectra. The obtained dataset is therefore similar to the raw spectra extracted from the HARPS-N ESPRESSO pipeline with the exception that they have been normalized using RASSINE \citep[][]{Cretignier:2020ab}. The results of our trained neural network can be seen in Fig.~\ref{Sun_diff}. Compared with the two other cases presented above, we find that the RMS of the RV, and BIS residual increase only slightly and signals associated with stellar activity can still be well modelled by the neural network. This is not surprising since the major YARARA correction besides the correction of stellar activity that improves the RVs is the correction of telluric lines. Given that we only use strong stellar lines that are not blended and affected by strong systematics to build our shape shells but also to compute the CCFs that are used to derive the RV time series series, we do not expect strong contamination from telluric features. For the FWHM though, the RMS of the residuals is however much larger, mainly induced by strong signals at one year and a half of it, due to the variation in the viewing angle of the Sun along the Earth orbit, but also a sudden jump in the FWHM time series around 59500 induced by the change of the HARPS-N cryostat.

\begin{figure*}[htbp]
\centering
\includegraphics[scale=0.4]{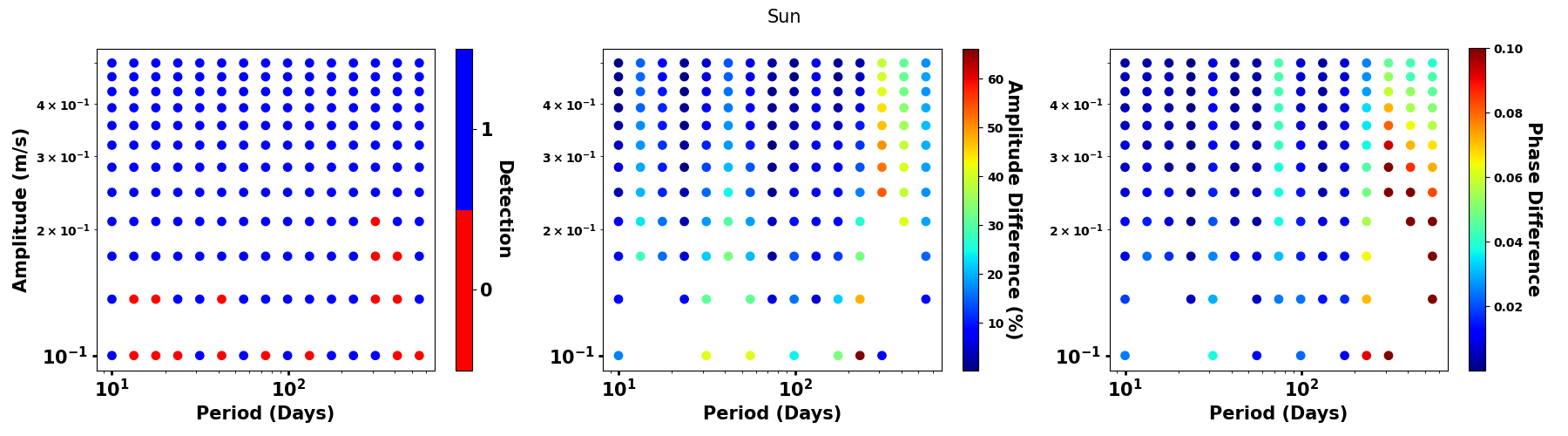}
\caption{Exoplanet detection limits from our neural network framework in the case of HARPS-N solar spectra. We derived the detection limit map by injecting simulated circular planetary signals covering periods ranging from 10 to 550 days and semi-amplitude ranging from 0.1 to 0.5 $\rm{m/s}$, into the solar data. \emph{Left:} Detection limit map in the period-amplitude domain. The red dots indicate the successful detection by the trained neural network with FAP $> 0.1\%$. \emph{Middle:} Amplitude comparison between the injected and recovered signals in the period-amplitude domain. The amplitude difference for most of the recovered signals are $< 20\%$. The large difference at low amplitude (0.1 m/s) is likely due to noise in the data. \emph{Right:} Phase comparison between the injected and recovered signals in the period-amplitude domain. The phase difference for most of the recovered planets is $< 0.04$. The large differences at long periods is likely due to a poorer sampling of the phase for long-period planetary signals.}
\label{Multi_kitcat_Detection_Limits_Sun}%
\end{figure*}

In order to test our neural network performance at recovering planetary signals, we injected simulated planets into the solar spectra. For the sake of simplicity, all the simulated planets have a circular orbit with amplitudes evenly sampled from $0.1$ to $0.5\,\rm{m/s}$ with a step size of $0.033\,\rm{m/s}$. The periods of the simulated planet signals are evenly sampled in logarithmic scale from 10 to 550 days, for a total of 180 planetary signals modeled. We inject each planetary signal into the solar spectral time series and train our neural network on that dataset. Since the $\log(R'_{HK})$ of the Sun does not change for each planet injection, there is no need to optimize again the neural network architecture.
We derive the RV residuals for each injection by subtracting the RVs predicted by the trained neural network from the RV time series which contains the planetary signal. We compute the periodogram of the RV residuals, with a corresponding numerically estimated $>0.1\%$ false alarm probability (FAP) level. A planet is detected if the power of its corresponding peaks is higher than the $>0.1\%$ FAP threshold, in which case we fit the observed signal with a sinus to recover the injected planet amplitude and phase. The results of this injection recovery test is illustrated in Fig.~\ref{Multi_kitcat_Detection_Limits_Sun}.

\begin{figure*}[htbp]
\centering
\includegraphics[scale=0.4]{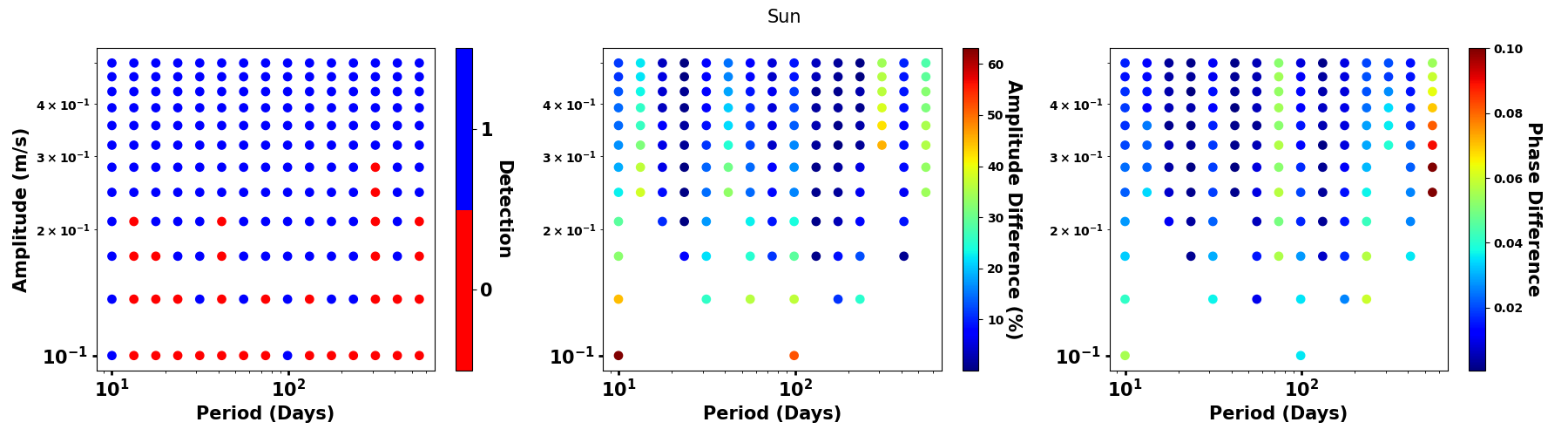}
\caption{Same as Fig.~\ref{Multi_kitcat_Detection_Limits_Sun} but with the YVA solar spectra.}
\label{Detection_Limits_Sun_matching_activity}%
\end{figure*}

\begin{figure*}[htbp]
\centering
\includegraphics[scale=0.4]{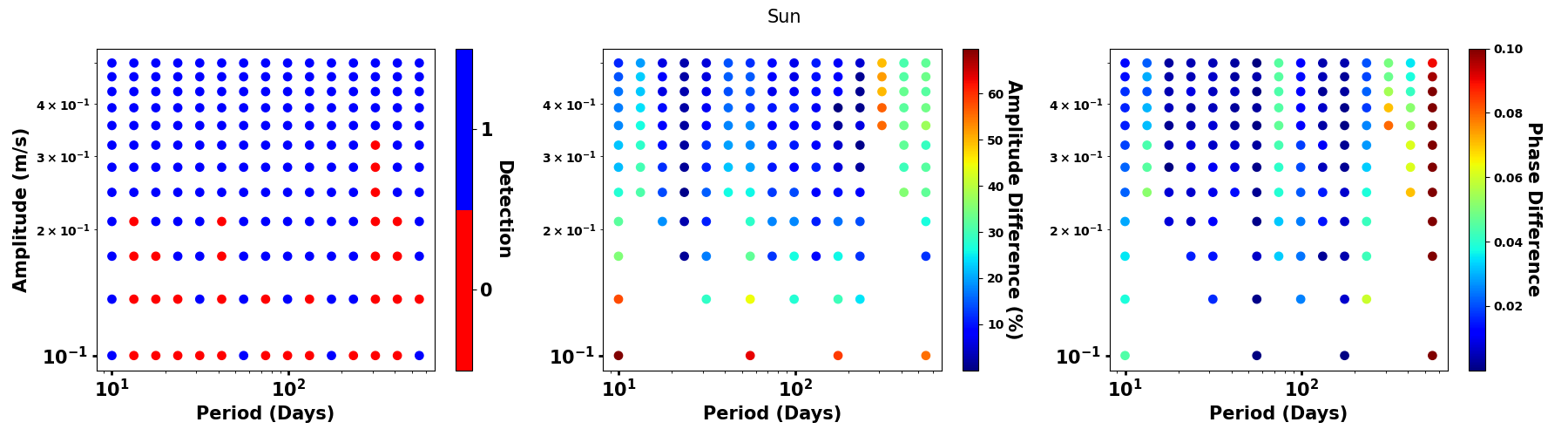}
\caption{Same as Fig.~\ref{Multi_kitcat_Detection_Limits_Sun} but with the YV0 solar spectra.}
\label{Detection_Limits_Sun_matching_diff}%
\end{figure*}

\begin{figure*}[htbp]
\centering
\includegraphics[scale=0.35]{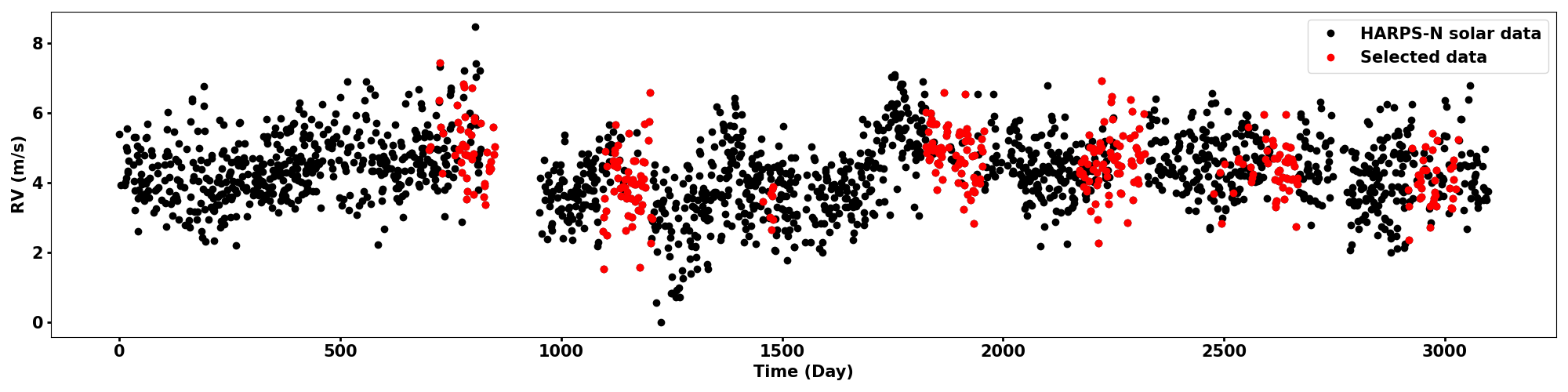}
\caption{Solar data were selected with the HD10700 sampling. We chose specific segments of the solar data sampled by HD10700 to evaluate the performance of the deep learning framework regarding the sampling effect. The full HARPS-N solar time series is labeled in black, while the selected data used for testing the sampling effect is labeled with red.}
\label{solar_HD10700_sampling}%
\end{figure*}

\begin{figure*}[htbp]
\centering
\includegraphics[scale=0.4]{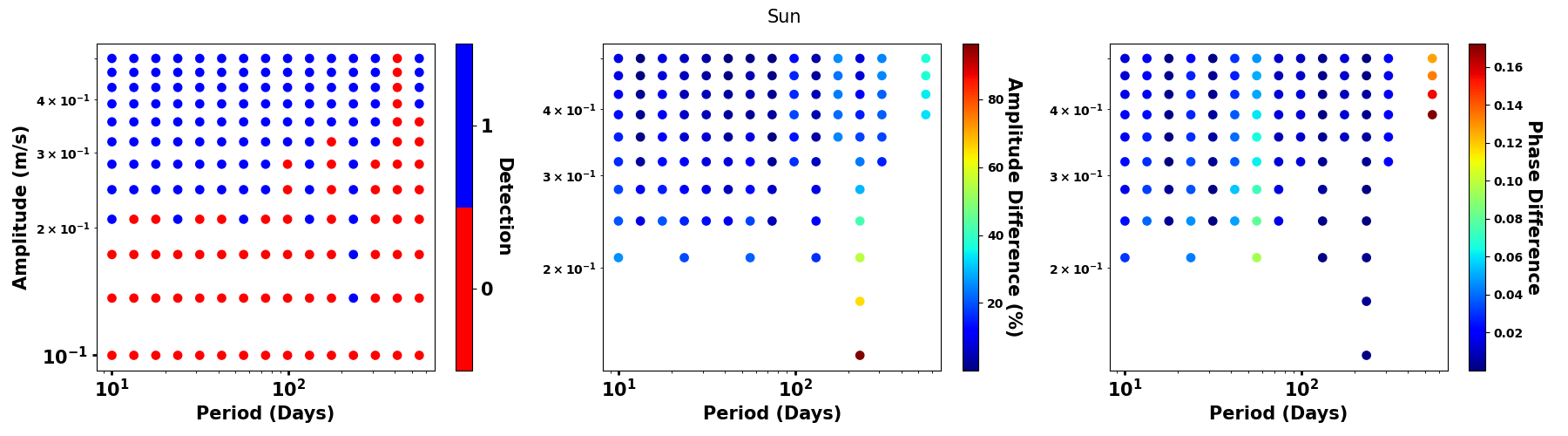}
\caption{Same as Fig \ref{Multi_kitcat_Detection_Limits_Sun} but for the HARPS-N solar spectral time series with the HD10700 sampling.}
\label{Multi_kitcat_Detection_Limits_sampling_HD10700}%
\end{figure*}

The trained neural network can reach a detection limit in planetary signal semi-amplitude as small as $0.18\,\rm{m/s}$ for period ranging from 10 to 550 days. For periods not close to the half period of the solar rotation and the 1 year observation window, the neural network can even reach a limit of $0.15\,\rm{m/s}$. Some of the successful detection at $0.1\,\rm{m/s}$ is very likely due to noise since their recovered amplitude and/or phase are significantly off from the injected values. The amplitude differences for most of the recovered planetary signals are less than $20\%$ and the phase differences less than $0.04$. We also test the effect of different YARARA corrections on the detection map. We also injected the simulated planet signals into the YVA solar spectra and the YV0 solar spectra. The detection maps corresponding to those different datasets are shown in Figs.~\ref{Detection_Limits_Sun_matching_activity} and \ref{Detection_Limits_Sun_matching_diff}, respectively. We found that the detection maps derived from those two different spectral datasets are similar to the one derived from the YV1 solar spectra except that the detection limits around half the solar rotation period and one year are higher than before. This is not surprising since we already found the YARARA correction does not improve a lot on the results when we applied the neural network to the solar spectra at different YARARA correction stages. While achieving a detection limit of $0.18\,\rm{m/s}$ for periods ranging from 10 to 550 days is indeed impressive, it represents an optimal scenario. This is because most stars do not have as frequent observations as the solar case. To further evaluate the performance of the deep learning framework regarding sampling effect. We selected segments of the HARPS-N spectral time series sampled at the observation timestamps of HD10700, as illustrated in Fig \ref{solar_HD10700_sampling}. We examined the detection thresholds of the neural network using the YV1 spectra with this sampling rate. The neural network demonstrates the capability to achieve a detection threshold of $\sim 0.35\,\rm{m/s}$ for planetary signals across periods spanning from 10 to 550 days, as illustrated in Fig \ref{Multi_kitcat_Detection_Limits_sampling_HD10700}.

\subsection{HD128621}

We used the HARPS spectra of HD128621 collected between 2008 February 28th and 2015 May 21st, which corresponds to 288 daily binned spectra with S/N of $\sim 600$. The best neural network architecture of HD128621 is summarized in Table \ref{tab:NN_HD128621}. The given neural network is trained with the cross-validation algorithm mentioned in \textbf{Algorithm \ref{algorithm1}} using the 288 spectral shape shell available. After 150 epochs of training, we derived the predicted RV, FHWM and BIS time series from the neural network outputs. The results are illustrated in Fig~.\ref{HD128621_mad}.

\begin{table}[H]
\centering
\caption{Neural network architecture for HD128621.}
\begin{tabular}{ c c c c c c}
                               \toprule

    \multicolumn{5}{c}{Input layer} \\ \midrule
    \multicolumn{5}{c}{Conv1 (112, 1)} \\
    \multicolumn{5}{c}{Conv2 (32, 3)} \\
    \multicolumn{5}{c}{Conv3 (80, 1)} \\ \midrule
    
    \multicolumn{5}{c}{Adaptive average pooling(3,3)} \\ \midrule
     
   FC1 (256) & \multicolumn{3}{c}{FC1 (256)} & FC1 (256) \\
   Dropout1 (0.2) & \multicolumn{3}{c}{Dropout1 (0.2)} & Dropout1 (0.2) \\

    FC1 (480) & \multicolumn{3}{c}{FC1 (480)} & FC1 (480) \\
   Dropout1 (0.2) & \multicolumn{3}{c}{Dropout1 (0.2)} & Dropout1 (0.2) \\

    FC1 (224) & \multicolumn{3}{c}{FC1 (224)} & FC1 (224) \\ \midrule
     
   FWHM & \multicolumn{3}{c}{RV} & BIS \\  \bottomrule
\end{tabular}
\tablefoot{Adam algorithm is employed as the optimizer.}
\label{tab:NN_HD128621}
\end{table}

We found that the RV time series of CCFs is significantly reduced from $2.166\,\rm{m/s}$ to $0.970\,\rm{m/s}$. The stellar activity signal at the rotation period of the star is well modeled, as shown in the RV periodogram. We see no significant signals left at the periodogram of the RV residuals. In the FWHM space, the trained neural network also generated a good fit to the FHWM time series derived from CCFs. In the BIS domain, the stellar activity feature in the periodogram can be captured by the neural network but there are still signals associated with stellar rotation left in the periodogram of the BIS residuals. For HD128621, we performed the same test as we did for the Sun. We checked how the results of the neural network change with different YARARA corrections. We trained the neural network on the YVA spectra of HD128621. The result is shown in Fig \ref{HD128621_activity}. Although the signals associated with stellar activity seem well modelled by the neural network in the periodogram of the RVs, FWHM and BIS, the RMS of the RV residual did increase a lot, from $0.970\,\rm{m/s}$ to $1.582\,\rm{m/s}$. Compared with the solar case, for which RV residual RMS only increases from $0.663\,\rm{m/s}$ to $0.872\,\rm{m/s}$, the trained neural network failed to model most of the components coming from the YARARA stellar activity correction. One possible explanation is that HD128621, which is a K-type star, has less stellar activity due to the inhibition of the convective blueshift than the solar case and strong flux effect \citep[e.g.]{Liebing-2021aa}. In order to further validate that the flux effect is the dominant component in the stellar activity of HD128621, we employed SOAP-GPU to evaluate the strength of different stellar activity components. We used the best-fit parameters of HD128621 from \cite{Dumusque-2014apj}. As shown in Fig.~\ref{activity_soap_gpu_HD128621}, the flux effect of HD128621 has an amplitude of $3.592\,\rm{m/s}$, which is larger than the amplitude of the convective blueshift inhibition. Details of the SOAP-GPU simulation on HD128621 can be found in Appendix \ref{sec_App_HD128621}. Given that our neural network framework is designed to model the effect induced by the convective blueshift inhibition, it is less effective at mitigating the flux effect component of stellar activity compared to the YARARA correction. We further tested the performance of the deep learning framework on the flux effect and the inhibition of the convective blueshift using SOAP-GPU simulated spectra. These spectra were generated based on the intensity continuum and magnetogram images obtained from the Helioseismic and Magnetic Imager (HMI/SDO; \cite{Schou-2012SoPh}). For further details regarding the results, please refer to Appendix \ref{sec_App_activity}.

\begin{figure*}[htbp]
\centering
\includegraphics[scale=0.4]{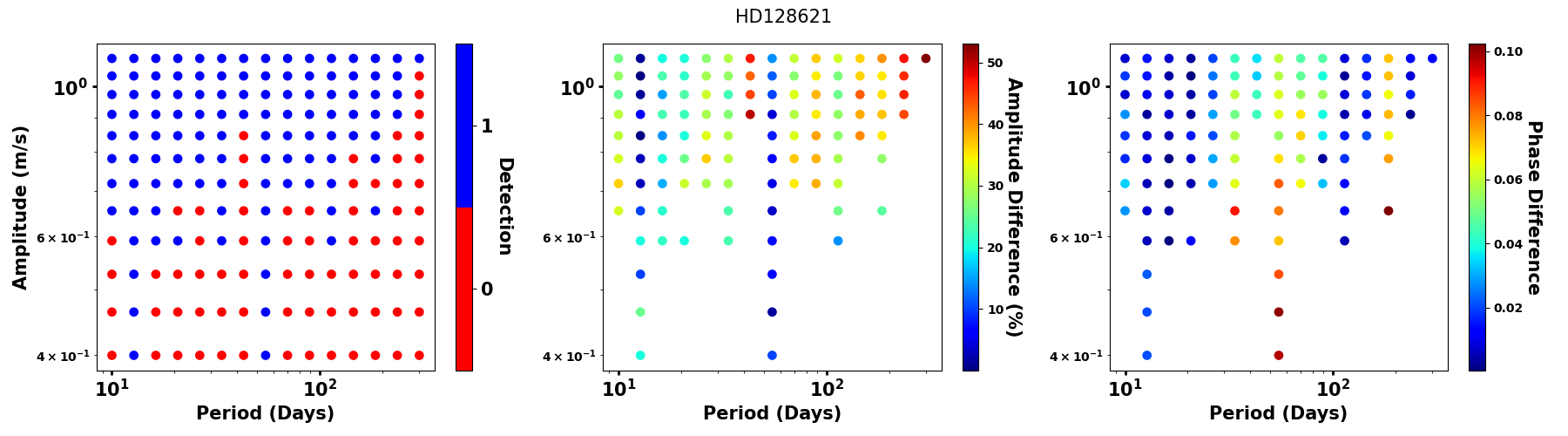}
\caption{Exoplanet detection limits on HD128621 HARPS spectra. We derived the detection limit maps by injecting simulated planetary signals covering periods ranging from 10 to 300 days and semi amplitude ranging from 0.4 to 1.1 $\rm{m/s}$, into the spectra. \emph{Left:} Detection limit map in the period-amplitude domain. The red dots indicate the successful detection by the trained neural network of a signal with a FAP $> 0.1\%$. \emph{Middle:} Amplitude comparison between the injected and recovered signals in the period-amplitude domain. The amplitude difference for most of the detected signals is $< 30\%$. \emph{Right:} Phase difference between the injected and recovered signals in the period-amplitude domain. The phase difference for most of the detected signals is $< 0.06$.}
\label{Detection_Limits_HD128621_matching_mad}%
\end{figure*}

\begin{figure*}[htbp]
\centering
\includegraphics[scale=0.4]{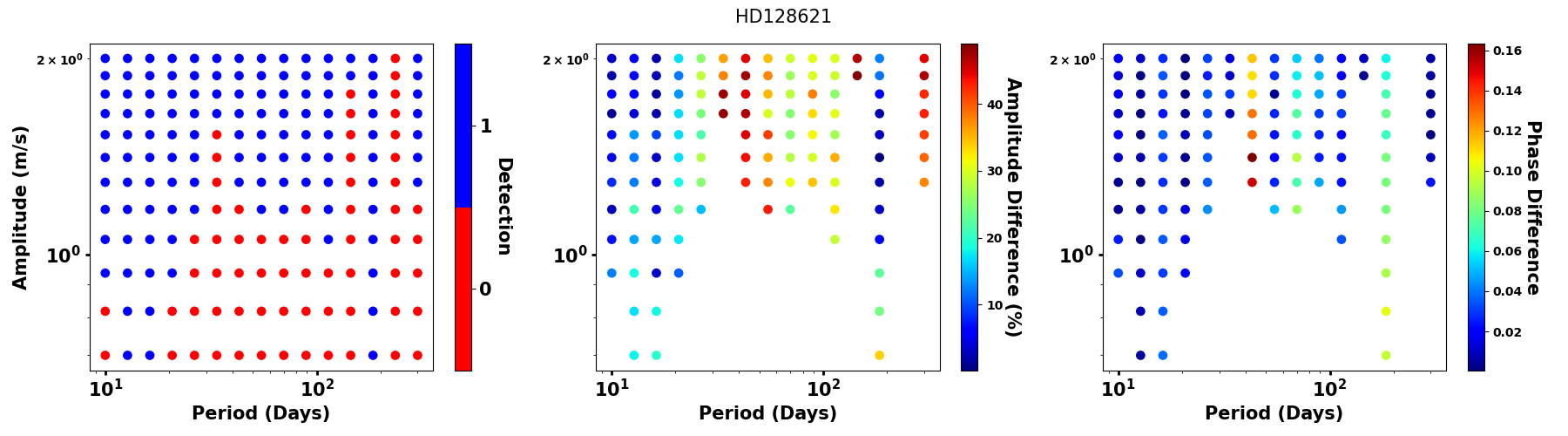}
\caption{Same as Fig \ref{Detection_Limits_HD128621_matching_mad} but with the YVA spectra of HD128621. Given that the neural network does not perform well on the YVA spectra, the semi amplitude of simulated planet signals is from 0.7 to 2.0 $\rm{m/s}$.}
\label{Detection_Limits_HD128621_matching_activity}%
\end{figure*}

We further test the neural network performance when an exoplanet signal is present in the spectral time series of HD128621. As in the solar case, a set of simulated circular planetary signals are created with the amplitude evenly sampled from $0.4\,\rm{m/s}$ to $1.1\,\rm{m/s}$ and periods evenly sampled in logarithmic scale from 10 days to 300 days, for a total of 180 simulated planetary signals. We trained our neural network on the datasets with the planetary signals injected at the spectral level. The derived detection limit map is shown in Fig \ref{Detection_Limits_HD128621_matching_mad}. We find that the detection limit in RV semi-amplitude of the trained neural network is $\sim 0.7\,\rm{m/s}$ for planetary signals with period ranging from 10 to 300 days. For signals with periods far from the stellar rotation period and half of is, the neural network can even reach a limit of $0.5\,\rm{m/s}$. The low detection limits around the stellar rotation period and half of it is due to the presence of remaining stellar activity at the RV residual level. We also test the effect of the YARARA stellar activity correction on the detection map by training the neural network on the YVA spectra (see Fig.~\ref{Detection_Limits_HD128621_matching_activity}. We find that the detection limit largely increases from $\sim 0.7\,\rm{m/s}$ to $\sim 1.7\,\rm{m/s}$. This is likely due to the flux effect component of active regions which is not well modeled in the spectral shell representation.

\subsection{HD10700} 

We used the HARPS spectra of HD10700 collected between 2005 August 1st to 2014 December 18th, which corresponds to 380 daily binned spectra with S/N of $\sim 500$. The best neural network architecture for HD10700 is summarized in Table \ref{tab:NN_HD10700}. Given the fact that HD10700 is a stellar quiet star and stellar activity may not be the dominant component, we trained the neural network on the 380 corresponding shape shells for 1000 training epochs. The predicted RV, FHWM and BIS time series from the neural network are shown in Fig \ref{HD10700_mad}.

\begin{figure*}[htbp]
\centering
\includegraphics[scale=0.4]{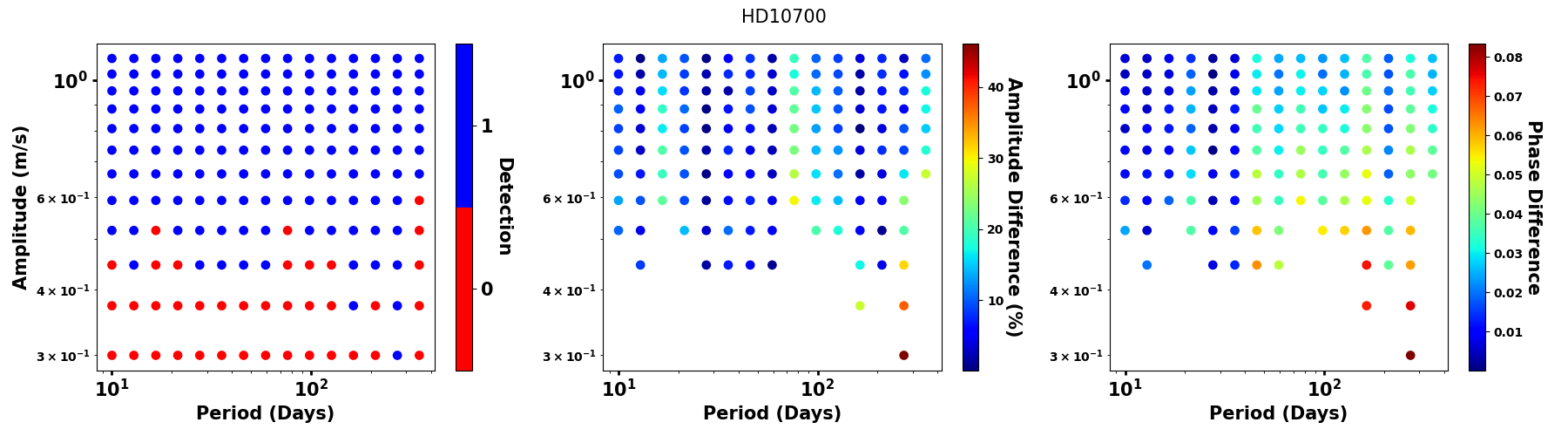}
\caption{Exoplanet detection limits on HD10700 HARPS spectra. We derived the detection limit maps by injecting simulated planetary signals covering periods ranging from 10 to 350 days and semi amplitude ranging from 0.3 to 1.1 $\rm{m/s}$, into the spectra. \emph{Left:} Detection limit map in the period-amplitude domain. The red dots indicate the successful detection by the trained neural network of a signal with a FAP $> 0.1\%$. \emph{Middle:} Amplitude comparison between the injected and recovered signals in the period-amplitude domain. The amplitude difference for most of the detected signals is $< 20\%$. \emph{Right:} Phase difference between the injected and recovered signals in the period-amplitude domain. The phase difference for most of the detected signals is $< 0.04$.}
\label{Multi_kitcat_Detection_Limits_HD10700}%
\end{figure*}

Because little stellar activity is present in the RV time series, the RMS of the RV residuals only decreases by $0.1\,\rm{m/s}$. No significant improvement in the FWHM and the BIS time series either. A summary of the neural network performance on HD10700, as well as on the Sun and HD128621, is provided in Table \ref{tab:summary}. Given the fact that instrumental systematic uncertainty is the dominant factor in the spectra of HD10700, it is difficult for the neural network to extract information of it at the level of the shape shells. We also tested the ability of our neural network to recover injected planetary signals in the spectra of HD10700. We simulated a set of circular planetary signals with the amplitude evenly sampled from $0.3\,\rm{m/s}$ to $1.1\,\rm{m/s}$ and periods evenly sampled in logarithmic scale from 10 days to 350 days, for a total of 180 simulated planetary signals. These planetary signals are injected into the spectra which are further used for the neural network training. The detection map of the trained neural network on the spectra of HD10700 can be found in Fig \ref{Multi_kitcat_Detection_Limits_HD10700}.  This analysis shows a detection threshold of $\sim 0.5\,\rm{m/s}$ for planetary signals with periods ranging from 10 to 350 days. The detections at long periods are likely to be false positives as the recovered amplitudes and phases are significantly different from the injected values.

\begin{table}[!htp]
\centering
\caption{Neural network architecture for HD10700.}
\begin{tabular}{ c c c c c c}
                               \toprule

    \multicolumn{5}{c}{Input layer} \\ \midrule
    \multicolumn{5}{c}{Conv1 (96, 5)} \\
    \multicolumn{5}{c}{Conv2 (48, 2)} \\  \midrule
    
    \multicolumn{5}{c}{Adaptive average pooling(3,3)} \\ \midrule
     
   FC1 (224) & \multicolumn{3}{c}{FC1 (224)} & FC1 (224) \\
   Dropout1 (0.3) & \multicolumn{3}{c}{Dropout1 (0.3)} & Dropout1 (0.3) \\

    FC1 (192) & \multicolumn{3}{c}{FC1 (192)} & FC1 (192) \\
   Dropout1 (0.3) & \multicolumn{3}{c}{Dropout1 (0.3)} & Dropout1 (0.3) \\

    FC1 (128) & \multicolumn{3}{c}{FC1 (128)} & FC1 (128) \\ \midrule
     
   FWHM & \multicolumn{3}{c}{RV} & BIS \\  \bottomrule
\end{tabular}
\tablefoot{Adam algorithm is employed as the optimizer.}
\label{tab:NN_HD10700}
\end{table}

\begin{table*}[t]
\centering
\caption{Summary of neural network performance on different stars at different YARARA corrections.}
\begin{tabular}{llllllllll}
\hline
\multicolumn{1}{c}{} & \multicolumn{3}{l}{The Sun}                & \multicolumn{3}{c}{HD128621}                & \multicolumn{3}{c}{HD10700} \\ \hline
\multicolumn{10}{c}{YV1 Spectra}                                                                                                              \\ \hline
                     & RV    & FWHM  & \multicolumn{1}{l|}{BIS}   & RV    & FWHM   & \multicolumn{1}{l|}{BIS}   & RV      & FWHM    & BIS     \\ \hline
$\sigma_{raw}$ (m/s)                & 0.822 & 0.682 & \multicolumn{1}{l|}{0.480} & 2.166 & 1.241  & \multicolumn{1}{l|}{1.328} & 1.100   & 1.018   & 0.478   \\
$\sigma_{residual}$ (m/s)             & 0.663 & 0.553 & \multicolumn{1}{l|}{0.314} & 0.970 & 0.703  & \multicolumn{1}{l|}{0.611} & 0.997   & 0.897   & 0.482   \\ \hline
\multicolumn{10}{c}{YVA spectra}                                                                                                              \\ \hline
                     & RV    & FWHM  & \multicolumn{1}{l|}{BIS}   & RV    & FWHM   & \multicolumn{1}{l|}{BIS}   &         &         &         \\ \hline
$\sigma_{raw}$ (m/s)           & 2.206 & 3.592 & \multicolumn{1}{l|}{1.028} & 2.621 & 14.718 & \multicolumn{1}{l|}{3.339} &         &         &         \\
$\sigma_{residual}$ (m/s)          & 0.872 & 0.584 & \multicolumn{1}{l|}{0.339} & 1.582 & 1.642  & \multicolumn{1}{l|}{0.780} &         &         &         \\ \hline
\multicolumn{10}{c}{YV0 Spectra}                                                                                                              \\ \hline
                     & RV    & FWHM  & \multicolumn{1}{l|}{BIS}   &       &        & \multicolumn{1}{l|}{}      &         &         &         \\ \hline
$\sigma_{raw}$ (m/s)            & 2.446 & 4.516 & \multicolumn{1}{l|}{1.084} &       &        & \multicolumn{1}{l|}{}      &         &         &         \\
$\sigma_{residual}$ (m/s)           & 0.898 & 0.908 & \multicolumn{1}{l|}{0.363} &       &        & \multicolumn{1}{l|}{}      &         &         &         \\ \hline
\end{tabular}
\tablefoot{$\sigma_{residual}$ is the residual after after extracting the neural network's prediction from the original time series.}
\label{tab:summary}
\end{table*}

\section{Conclusions}\label{sec5}

In this paper, we developed a neural network framework to efficiently mitigate stellar activity at the spectral level, to enhance the detection of low-mass planets on periods from a few days up to a few hundred days, corresponding to the habitable zone of solar-type stars. The framework is flexible and can be adapted to any solar-like star, by optimizing the neural network architecture in a data-driven way, by using the $Optuna$ software and predicting the $\log(R'_{HK})$ index of a given star, an index that is not affected by planetary signals. The neural network takes spectral shape shell as the input to predict the corresponding RV, FWHM, and BIS. 

We tested this framework on three stars that have been intensively observed by HARPS and HARPS-N: the Sun, HD128621 and HD10700. In the solar case, the neural network is able to detect planet signals with semi amplitude as low as $0.2\,\rm{m/s}$ and periods ranging from 10 to 550 days. When applying the framework to the more active star HD128621, the neural network can reach a detection limit of $0.7\,\rm{m/s}$ for planet signals with periods from 10 to 300 days. In the case of the quiet star HD10700, the neural network does not help much in decreasing the RMS of the RV residuals, however, we can still reach a very low detection limit of $0.5\,\rm{m/s}$ in planetary semi-amplitude, for planets with periods ranging from 10 to 350 days.

We further tested how the different stages of YARARA corrections affect our neural network performances. We find that the YARARA stellar activity correction is important when likely the flux component of stellar activity is more important, like it should be the case for later spectral-type stars where the velocity of convection diminishes, like for the K-dwarfs HD128621. However, on a star like the Sun that is dominated by the stellar activity effect induced by the inhibition of the convective blueshift, for which shell have been designed to be sensitive for, our neural network framework seems to model efficiently stellar activity signals. 

To improve further the performance of our neural network framework, we should consider adding as input some information about the flux effect induced by stellar activity. This could be done either by injecting at the convolution layer input a spectral representation, similar to the shell, but which is optimized to extract the flux effect, or injecting at the level of the fully connect layer, time series that are sensitive to the flux effect, like photometry. We could further enhance our deep learning framework's performance by training an autoencoder to model shape shells for specific spectral types, rather than individual stars. Once the model is well-trained, we can use transfer learning to connect the pretrained encoder to new fully connected layers. This enables us to effectively train the neural network to model RV time series due to stellar activity.

The obtained results with our neural network approach for the Sun are quite impressive and to the best of our knowledge, it is the first time such a low detection threshold, $0.2\,\rm{m/s}$, can be reached on the HARPS-N solar dataset. Clearly, there are many possible ways to improve the proposed framework, and therefore, there is hope to reach one day the $0.1\,\rm{m/s}$ level, that would correspond to the detection of an Earth-like planet orbiting in the habitable zone of a solar-type star. In this study, we only consider the scenario where we blindly search for planetary signals in the RV time series. If planetary information such as transit time and phase are known, in addition to space-based photometry, what the PLATO mission \citep[][]{Rauer-2014} should deliver, it is possible that we could be sensitive to other Earths.

\begin{acknowledgements}
We thank the referee Luis Agustin Nieto for his valuable and constructive feedback. This project has received funding from the European Research Council (ERC) under the European Union’s Horizon 2020 research and innovation programme (grant agreement SCORE No 851555 and from the Swiss National Science Foundation under the grant SPECTRE (No 200021\_215200). This work has been carried out within the framework of the NCCR PlanetS supported by the Swiss National Science Foundation under grants 51NF40\_182901 and 51NF40\_205606. FPE would like to acknowledge the Swiss National Science Foundation (SNSF) for supporting research with HARPS-N through the SNSF grants nr. 140649, 152721, 166227, 184618 and 215190. The HARPS-N Instrument Project was partially funded through the Swiss ESA-PRODEX Programme.
\end{acknowledgements}

\bibliographystyle{aa}
\bibliography{Yinan_Zhao_bibli}

\begin{appendix}

\section{Results of the deep learning framework on different stars}\label{sec_App}

We summarize the results of the trained neural network on the Sun, HD128621 and HD10700. The predicted RV, FWHM and BIS time series of each case are derived by training the corresponding neural networks using the cross validation technique described in \textbf{Algorithm \ref{algorithm1}}. For each iteration in \textbf{Algorithm \ref{algorithm1}}, we randomly select $90 \%$ of the spectra as the training set and the rest $10 \%$ of the spectra as the test set. After the neural network is trained, the trained model is applied to remaining $10 \%$ of the spectra to get the predicted values. For the next iteration, $10 \%$ of the spectra, which are different from the last iteration, are selected as the test set and the remaining $90 \%$ are used for training. By repeating this process 10 times, all the spectra can be predicted by the same neural network architecture but with 10 sets of parameters.

\begin{figure*}[htbp]
\centering
\includegraphics[scale=0.35]{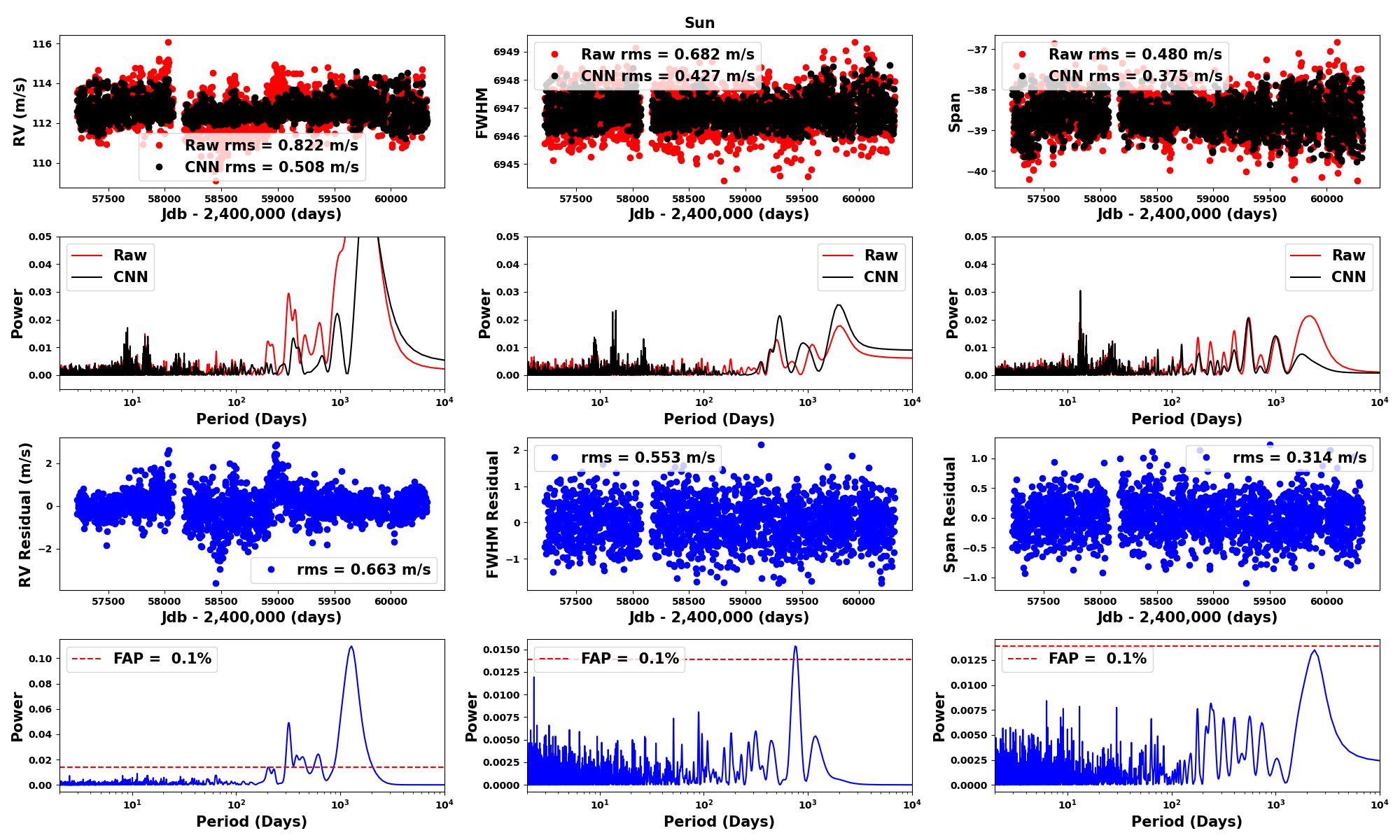}
\caption{Results of stellar activity modeling on the HARPS-N solar data with our trained neural network. The results of our neural network modeling on RVs, FWHM and BIS are shown in the left, middle and right columns, respectively. The first panels of each column shows the physical parameter derived from the specific solar cross-correlation mask and the one predicted by the neural network. Their corresponding periodograms are shown in the second panels of each column. The residuals of each physical parameter are shown in the third panels of each column. The last panels of each column shows the periodogram of the corresponding residuals. The FAP level of $0.1 \%$ is present in red dashed line.}
\label{Sun_mad}%
\end{figure*}

\begin{figure*}[htbp]
\centering
\includegraphics[scale=0.35]{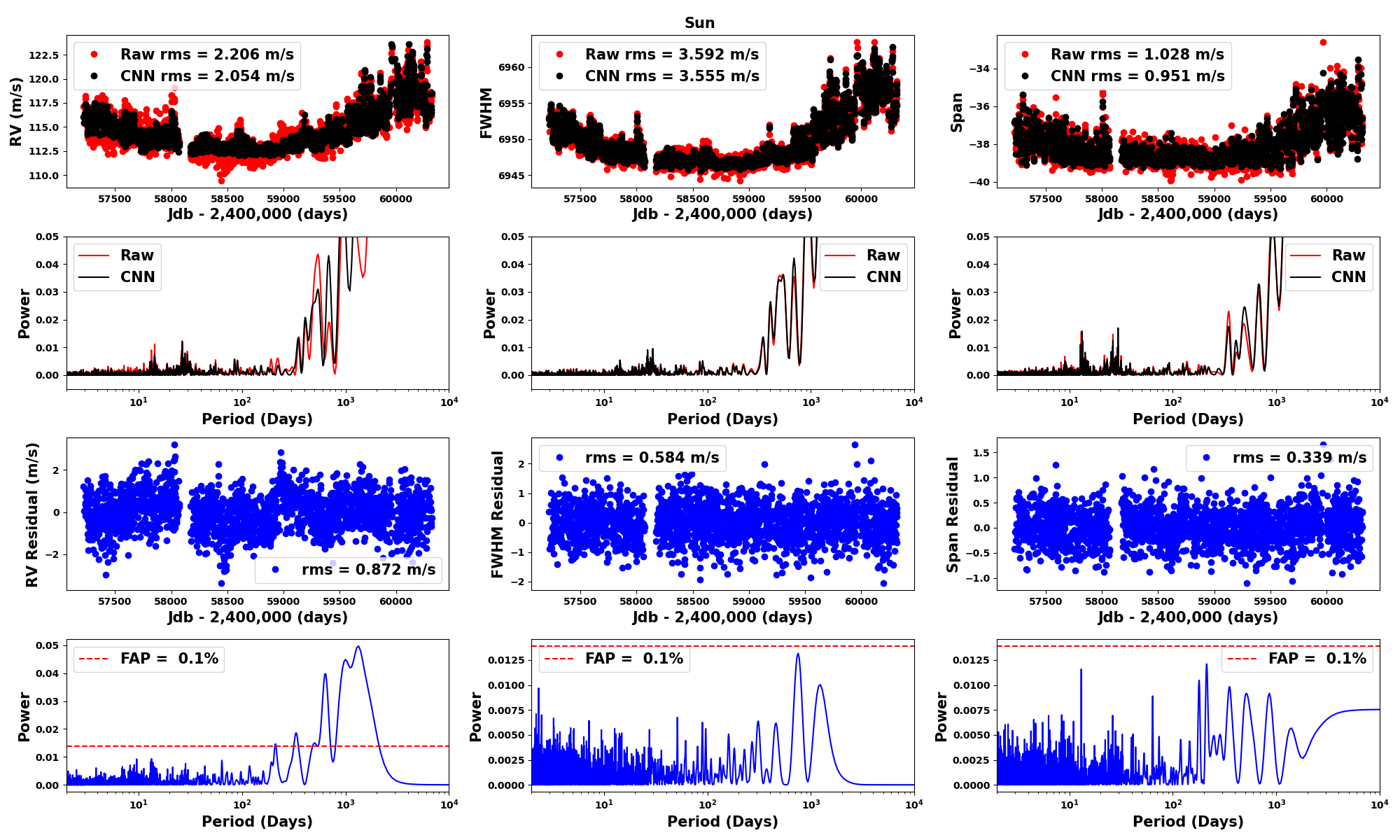}
\caption{Same as Fig \ref{Sun_mad} but with the YVA solar spectra.}
\label{Sun_activity}%
\end{figure*}

\begin{figure*}[htbp]
\centering
\includegraphics[scale=0.35]{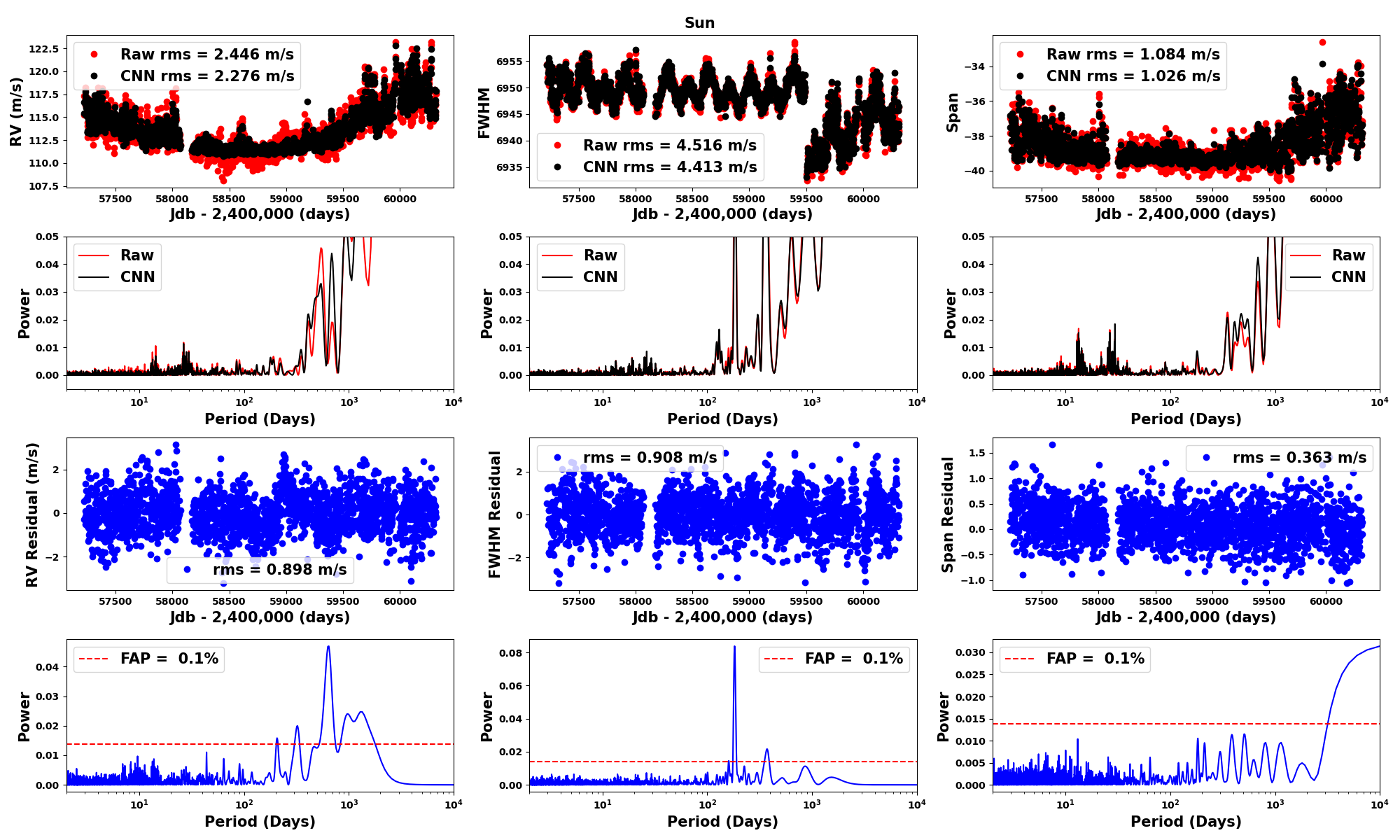}
\caption{Same as Fig \ref{Sun_mad} but with the YV0 solar spectra.}
\label{Sun_diff}%
\end{figure*}

\begin{figure*}[htbp]
\centering
\includegraphics[scale=0.35]{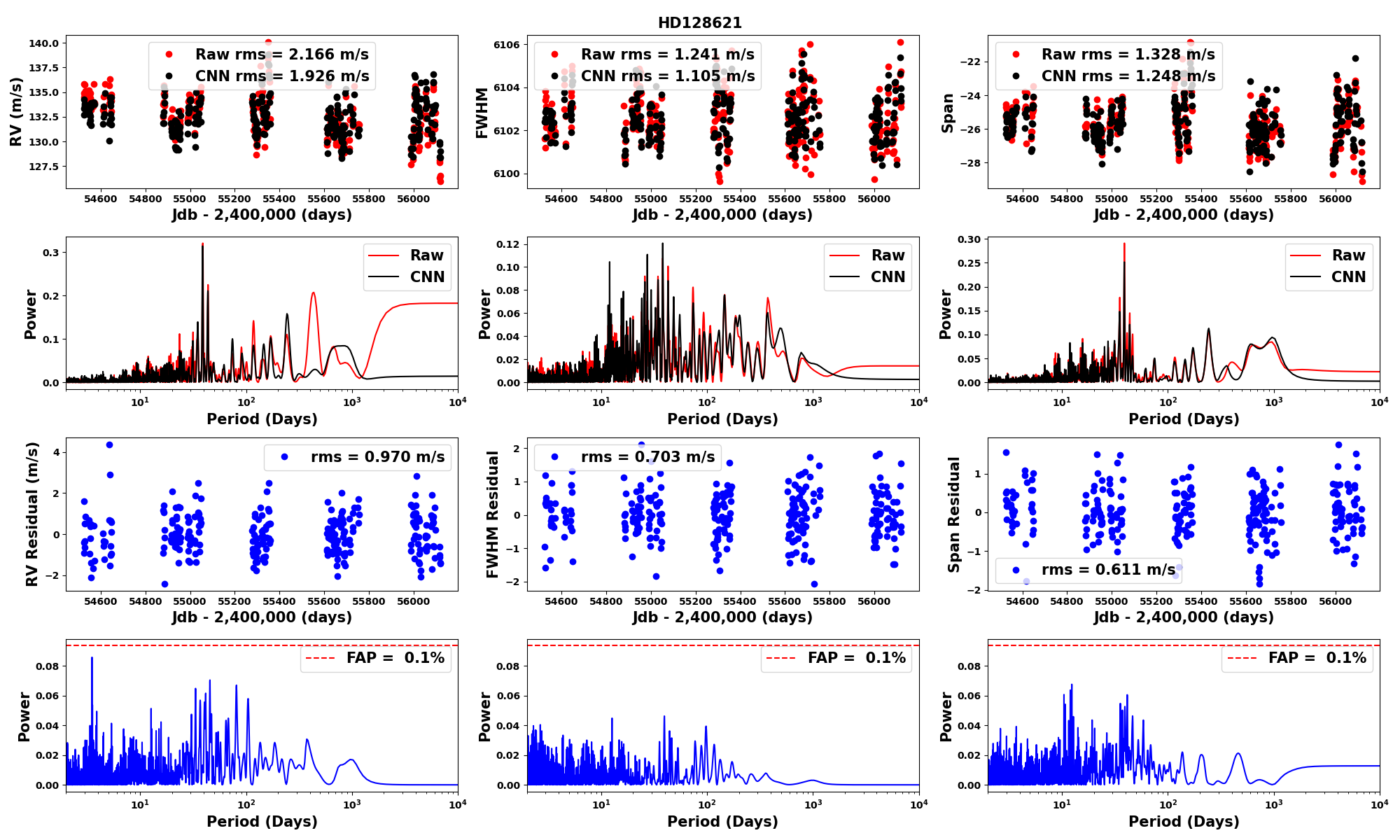}
\caption{Same as Fig \ref{Sun_mad} but for the HARPS data of HD128621.}
\label{HD128621_mad}%
\end{figure*}

\begin{figure*}[htbp]
\centering
\includegraphics[scale=0.35]{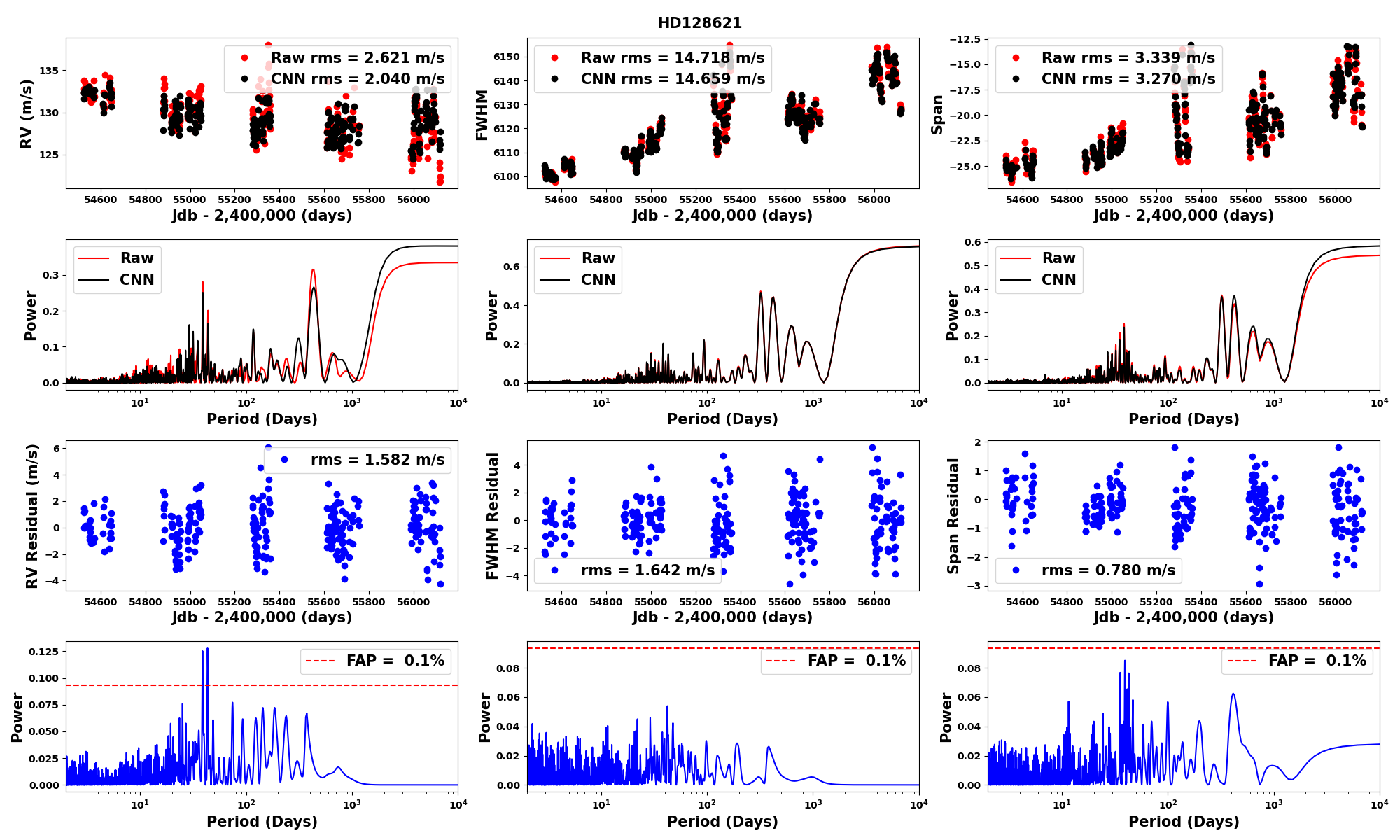}
\caption{Same as Fig \ref{HD128621_mad} but with the YVA spectra of HD128621.}
\label{HD128621_activity}%
\end{figure*}

\begin{figure*}[htbp]
\centering
\includegraphics[scale=0.35]{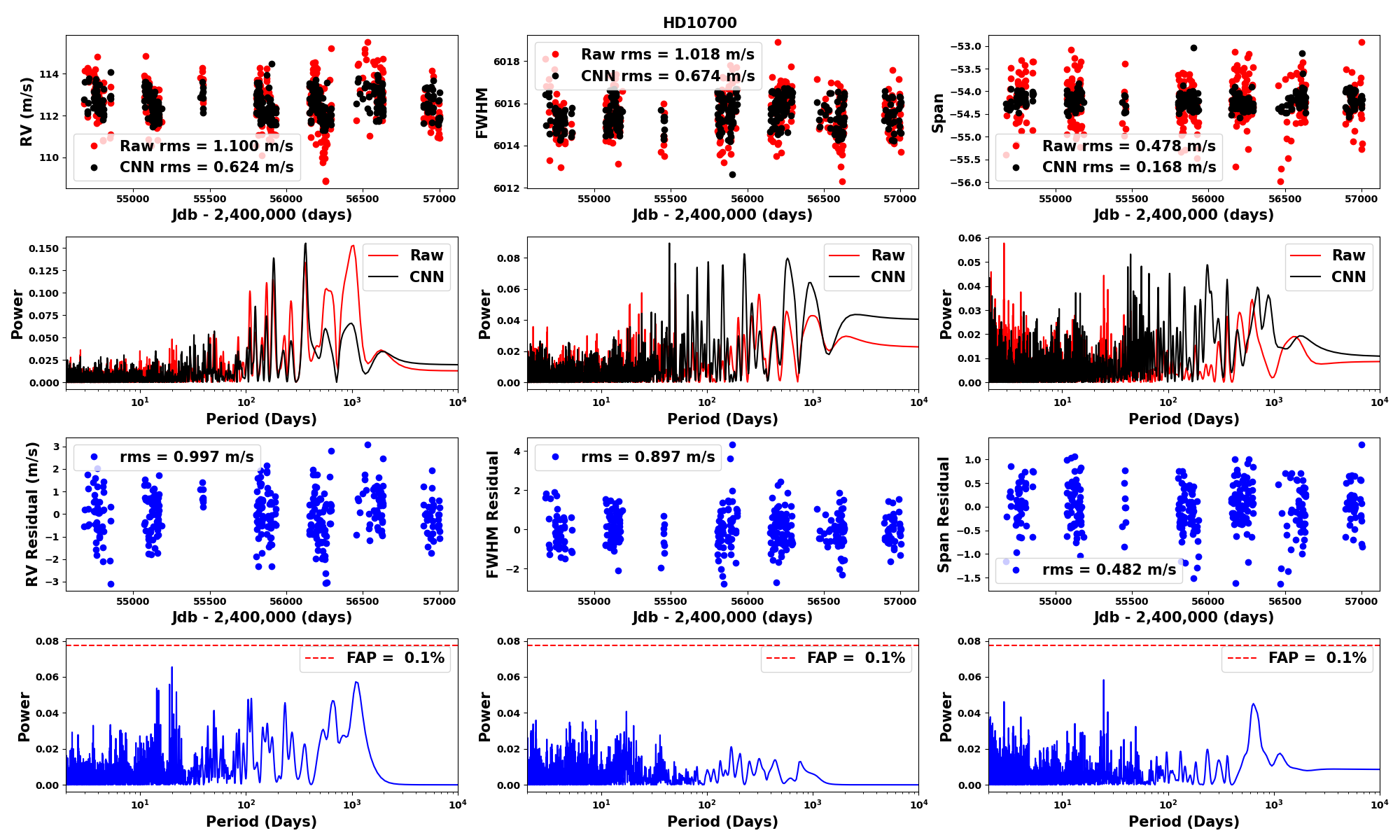}
\caption{Same as Fig \ref{Sun_mad} but for the HARPS data of HD10700.}
\label{HD10700_mad}%
\end{figure*}

\section{Stellar activity contributions of HD128621}\label{sec_App_HD128621}

We employed SOAP-GPU to evaluate the importance of flux effect and the inhibition of convective blueshift in HD128621. We adopted the best-fit parameters of HD128621 from \cite{Dumusque-2014apj}: A single faculae region with a size of $2.4 \%$ at latitude of 44 degrees, and the host star has a rotation period of 36.65 days with an inclination of 22 degrees. Using these parameters, we found that the amplitude of the flux effect is $3.592\,\rm{m/s}$ and the amplitude of the convective blueshift inhibition is $3.053\,\rm{m/s}$, which is smaller than the flux effect. The simulation further validates that the flux effect is the dominant component in the stellar activity of HD128621.

\begin{figure}[htbp]
\centering
\includegraphics[scale=0.35]{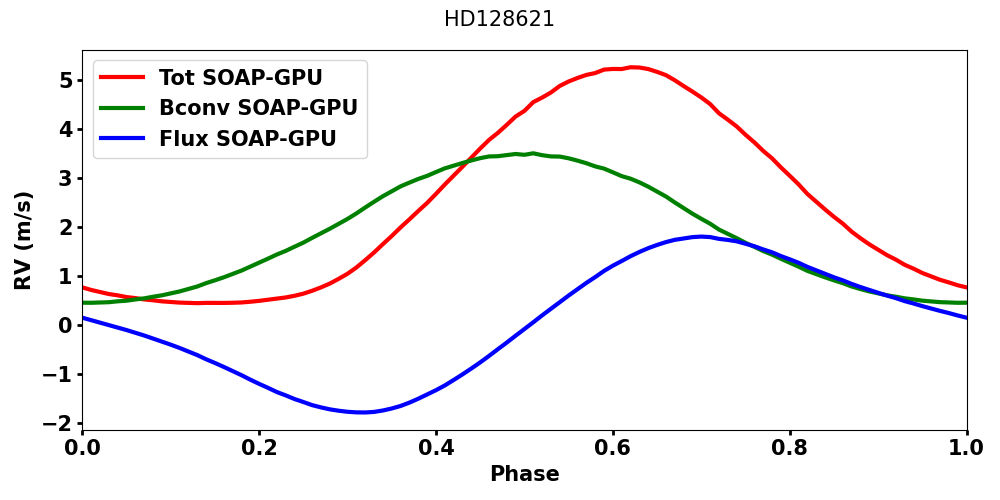}
\caption{Different stellar activity components of HD128621 simulated by SOAP-GPU. The flux effect, inhibition of convective blueshift, and the total effect are labeled with blue, green, and red, respectively. The amplitude of the flux effect is $3.592\,\rm{m/s}$ while the inhibition of convective blueshift has the amplitude of $3.053\,\rm{m/s}$. The amplitude of the total effect is $4.818\,\rm{m/s}$.}
\label{activity_soap_gpu_HD128621}%
\end{figure}

\section{Results of the deep learning framework on different stellar activity effects}\label{sec_App_activity}

\begin{figure*}[htbp]
\centering
\includegraphics[scale=0.35]{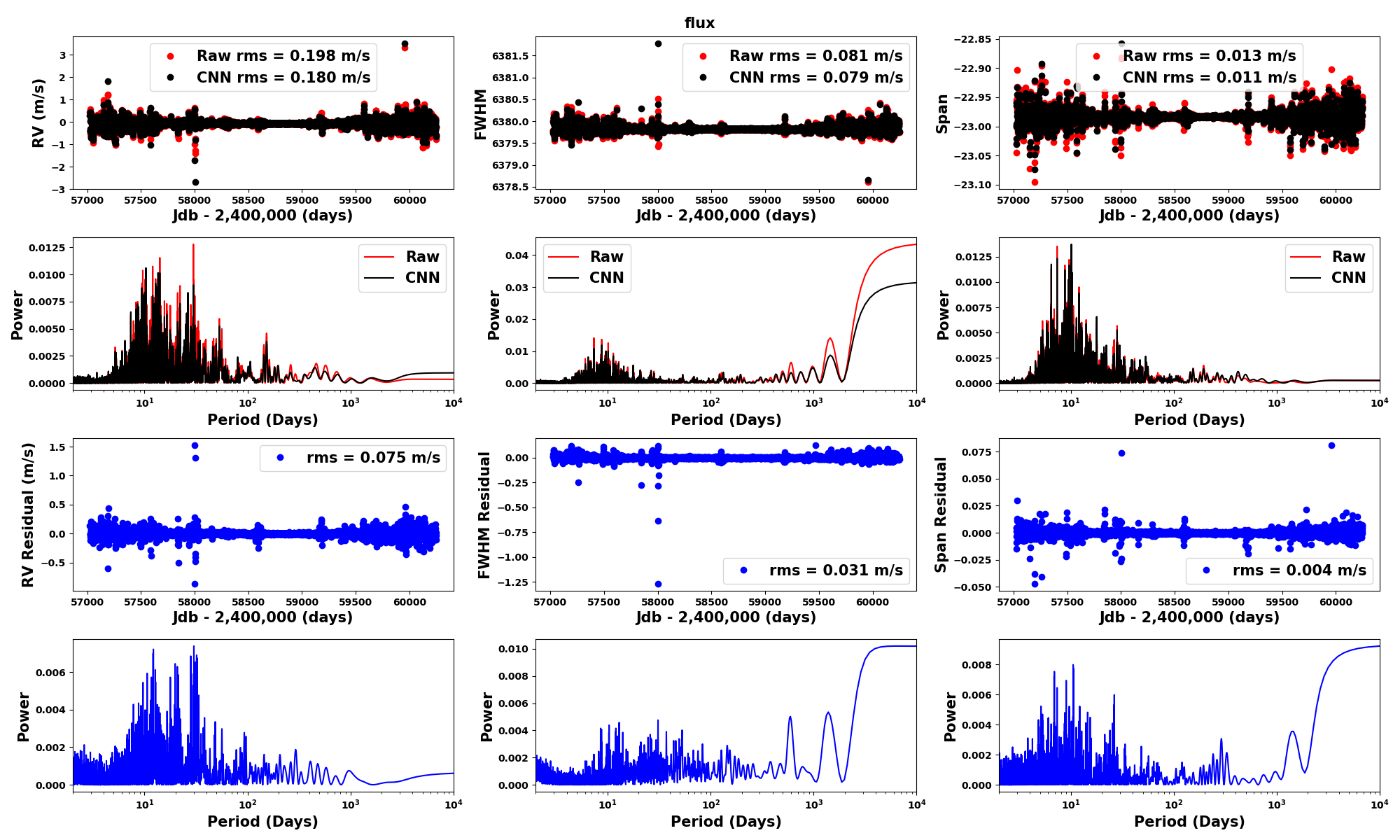}
\caption{Same as Fig \ref{Sun_mad} but for the SOAP-GPU simulated spectra of the flux effect.}
\label{cnn_flux}%
\end{figure*}

\begin{figure*}[htbp]
\centering
\includegraphics[scale=0.35]{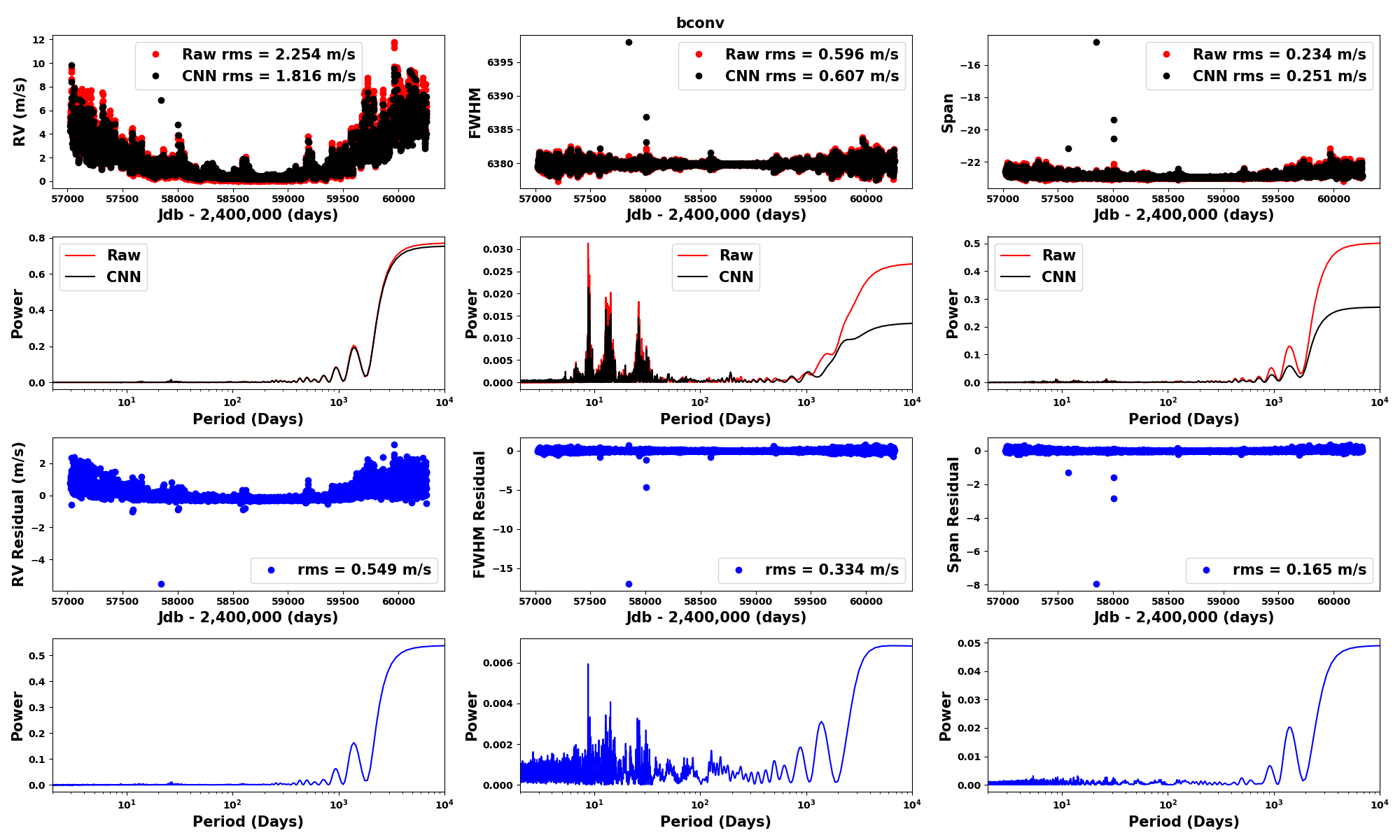}
\caption{Same as Fig \ref{Sun_mad} but for the SOAP-GPU simulated spectra of the inhibition of the convective blueshift effect.}
\label{cnn_bconv}%
\end{figure*}

The two primary effects of stellar activity are the flux effect and the inhibition of the convective blueshift. We employed the deep learning framework on simulated spectra generated by SOAP-GPU to evaluate the neural network's performance on each effect. These simulated spectra were derived from SDO/HMI images, which effectively model solar activity (Zhao et al. in prep). The neural network's results for each effect are illustrated in Fig \ref{cnn_flux} and Fig \ref{cnn_bconv}, respectively. Our findings reveal that the RMS of the RV time series can be reduced by 62 $\%$ for the flux effect, and by 76 $\%$ for the inhibition of the convective blueshift. This suggests that the deep learning framework is particularly effective in correcting the inhibition of the convective blueshift in shell space.

\end{appendix}
\end{document}